\documentclass[aps,pra,twocolumn,superscriptaddress,longbibliography]{revtex4-2}

\usepackage{amsmath,amssymb}
\usepackage{graphicx}
\usepackage{xcolor}
\usepackage{braket}
\usepackage{algorithm}
\usepackage{algorithmic}
\usepackage{url}
\usepackage{hyperref}
\usepackage{placeins}

\graphicspath{{figures/}}
\newcommand{\rev}[1]{#1}

\begin{document}

\title{Global Clock-Based Stochastic Modeling of a Memory-Assisted Measurement-Device-Independent Quantum Key Distribution System}

\author{Fares Nada}
\affiliation{Department of Electronics, Carleton University, Ottawa, Ontario K1S 5B6, Canada}

\author{Amber Hussain}
\affiliation{Department of Electronics, Carleton University, Ottawa, Ontario K1S 5B6, Canada}

\author{Tasmi Ahmed}
\affiliation{Department of Electronics, Carleton University, Ottawa, Ontario K1S 5B6, Canada}

\author{Andrew MacRae}
\affiliation{Department of Physics and Astronomy, University of Victoria, Victoria, BC V8W 2Y2, Canada}

\author{Connor Kupchak}
\email{connorkupchak@cunet.carleton.ca}
\affiliation{Department of Electronics, Carleton University, Ottawa, Ontario K1S 5B6, Canada}

\begin{abstract}
On the pathway to quantum key distribution on a global scale will be the realization of metropolitan-sized, Memory-Assisted Measurement-Device-Independent Quantum Key Distribution (MA-MDI-QKD) systems. Here, we present a simplistic and intuitive stochastic model to predict error-corrected key throughput in a MA-MDI-QKD scheme that incorporates a series of channel and quantum memory parameters. Building on our recently introduced global and local clock framework for entanglement distribution, our model employs a global clock system that accounts for the asynchronous arrival of Alice and Bob's qubits at the central station through leveraging quantum memories.  By focusing on metropolitan distances, we predict an error-corrected throughput metric over ranges of 10-50 km.  Our streamlined approach is a valuable addition to existing quantum network simulators that will guide the smooth integration of quantum networking into the field of communications engineering.
\end{abstract}


\maketitle

\section{Introduction}\label{Sec:1}

Over the past decades, quantum key distribution (QKD) has been recognized as a leading solution to exchange cryptographic keys with information-theoretic security under the assumptions of the protocol. From its infancy, BB84 \cite{Bennett1984} has been celebrated as a provably secure protocol for distributing symmetric encryption keys. However, BB84 encounters a number of practical challenges that can hinder its real-world application, including detector imperfections, the need for stringent quantum light sources and the degradation of key rate with optical loss \cite{PracticalQKD2020}.

More recently, an enhanced quantum key distribution scheme was introduced that is intrinsically secure against detector side-channel attacks, known as measurement-device-independent QKD (MDI-QKD) \cite{LoMDIQKD2012,Valivarthi2015}. The concept of MDI-QKD stems from performing a time-reversed version of entanglement-based QKD \cite{E91} and requires a Bell state measurement (BSM) at an intermediate node between parties (conventionally named Charlie) to signal correlations between Alice and Bob. In this scheme, the secrecy of the key is robust even when Charlie and the measurement apparatus cannot be trusted, providing enhanced security compared to traditional QKD \cite{LoMDIQKD2012,Rubenok2013}.

The establishment of globally connected QKD networks will begin at metropolitan distances \cite{BostonQuantumNetwork,pirandola2015}, where quantum communications have now been demonstrated in real-world settings \cite{Valivarthi2016,Craddock2024}, including MDI-QKD \cite{JianWeiPan2016MDIQKD,JianWeiPan2020} and very recently MA-MDI-QKD \cite{Bhaskar2020}. Although many implementations use fibre-based networks \cite{BostonQKD2018}, free-space and satellite links are also being pursued \cite{Sit2017,Pugh2017,Brougham2022,Ghalaii2023} and introduce challenges such as atmospheric turbulence \cite{Jaouni2025}, adaptive optics \cite{Scarfe2025} and beam tracking \cite{Pugh2017}.

{Several approaches are being pursued to overcome the rate-distance limitation of point-to-point QKD. Twin-field QKD, for example, has demonstrated secure key generation over more than 830~km of optical fibre \cite{WangTF2022}. Memory-assisted schemes offer a complementary route that can specifically be aimed at metropolitan multi-user networks, where quantum memories at an untrusted central node can buffer independent channel losses. The present work addresses this memory-assisted, network-engineering route and does not compete directly with point-to-point long-distance QKD.\par}

Predictive models for city-scale MDI-QKD schemes will be vital as these networks become established in our communication infrastructures. To date, these types of MDI-QKD models have seen limited investigation  \cite{Wang2014Simulating}. Here, we present a stochastic modeling algorithm based on Monte Carlo sampling that allows the prediction of error-corrected key throughput using a MA-MDI-QKD system over metropolitan distances. Our model analyzes an error-centric key rate with separation of the physical layer from the cryptography layer to evaluate the bit-error-limited throughput of a classical optical link before adding authentication and encryption overhead, where one can potentially first optimize memory before layering decoy constraints in a second stage. We believe this model augments the current family of quantum network simulators~\cite{Bel2025}.

\section{Memory-assisted MDI-QKD}\label{Sec:2}
Extending the MDI-QKD protocol to include quantum memories retains the original benefits of the protocol while enabling asynchronous arrival of Alice and Bob's qubits at the central node. The scaling of the transmission $\eta_t$ then improves from approximately $\sim\mathcal{O}(\eta_t(L))$ to $\sim\mathcal{O}(\sqrt{\eta_t(L)})$, significantly enhancing distribution rates over long distances. 

\begin{figure*}[!t]
    \centering
    \includegraphics[width=0.85\textwidth,height=0.42\textheight,keepaspectratio]{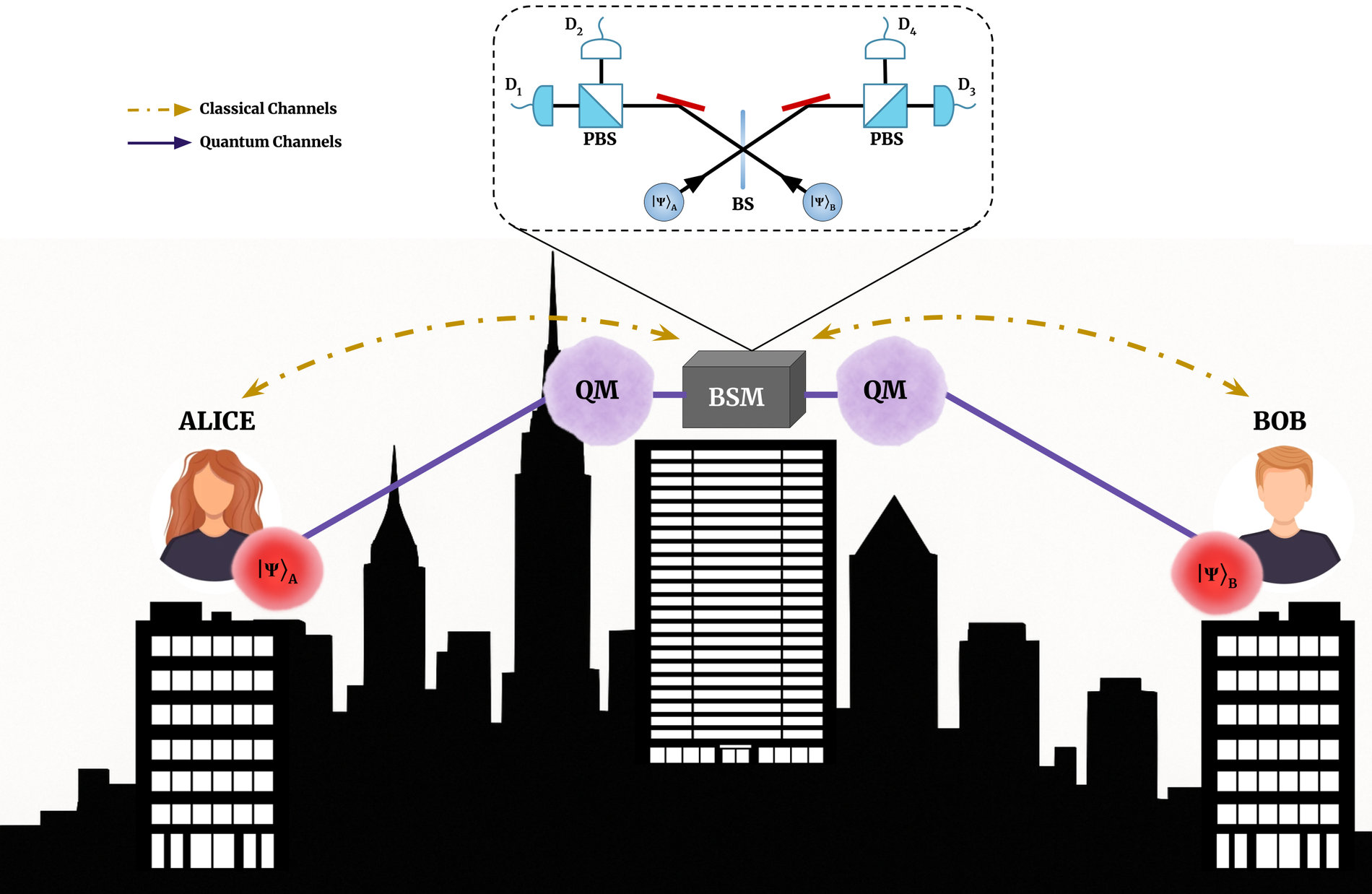}
    \caption{Schematic of intra-city quantum communications using polarization-based MA-MDI-QKD. Alice and Bob hold qubit states at their stations situated on metropolitan rooftops. Charlie's untrusted central node is on an intermediate rooftop hosting a polarization-based BSM apparatus consisting of a beamsplitter (BS), two polarization beam splitters (PBS) and four single-photon detectors $D_{1-4}$. Quantum channels are for qubit transmission and classical channels are for exchanging basis choices and notification of a successful BSM.}
    \label{CitySchem}
\end{figure*}

As illustrated in Fig.~\ref{CitySchem}, MA-MDI-QKD shares similarities with a single-node quantum repeater.  Alice and Bob hold emissive sources and Charlie hosts an intermediate station with quantum memories present for each channel \cite{Panayi2014,Simon2011}, unlike alternate protocols that place entanglement-generating memories along the link and use swapping to extend reach. In our model, the memories buffer the independently transmitted qubits until both memories have been loaded.  The failed loading and memory-induced noise explicitly enter into the rate equations from the gain and QBER modeling. The achievable error-corrected throughput is reliant on memory efficiencies and the indistinguishability of the photonic modes over background noise. Charlie and the measurement apparatus may be fully untrusted as the raw key derives from two-photon interference correlations that do not reveal prepared states.

The information-based steps of the MA-MDI-QKD protocol with polarization qubits are as follows:  Alice and Bob each randomly prepare and record one of four BB84 polarization states, ${\ket{H},\ket{V},\ket{+},\ket{-}}$, in the Z-basis ${\ket{H},\ket{V}}$ or X-basis ${\ket{+},\ket{-}}$ \cite{Bennett1984}, then transmit each qubit to Charlie's intermediate station where a dedicated QM loads the channel. Once both QMs herald a successful load, the stored states are retrieved for a BSM at Charlie, who announces the projection outcome to Alice and Bob so they update or discard their recorded bits or restart if the measurement fails. When sufficient successful events are collected, Alice and Bob announce their basis choices over a classical channel (Fig.~\ref{CitySchem}) and retain only matching-basis bits to form a shared raw key. The raw key undergoes standard BB84 post-processing with error correction and privacy amplification to yield a final secret key.

We set out for our model to simulate MA-MDI-QKD distribution rates at urban distances \cite{AdvancedQKD2024,Krzic2023,Zhang2025} such as high-rise rooftops like that shown in Fig. \ref{CitySchem}.  Charlie's central node is located between the two parties, containing a polarization-based BSM composed of a 50/50 beam splitter, two polarization beam splitters, and four single-photon detectors.  Our model will incorporate channel-rated losses and stochastically model quantum memory effects.

\section{Stochastic modeling of MA-MDI-QKD systems}\label{Sec:3}

Predictive models have been developed for BB84 and entanglement-based systems \cite{Scarani2009}, including stochastic software packages for BB84 \cite{Jasim2015,Kohnle2017} and entanglement-based distribution networks \cite{Wu2021,Semenenko2022,Ahmed2026}, but reports on MDI-QKD simulation methods remain limited \cite{Simon2018,Wang2014Simulating}. 

Stochastic modeling is naturally suited for memory-based schemes, as QMs involve many probabilistic processes occurring simultaneously. These include photon arrival times (often Poisson-distributed), memory loading/retrieval success (efficiencies < 1), decoherence over time, entanglement swapping success, loss, dark counts, timing jitter, etc. These processes are also required to be correlated as a late-arriving qubit affects memory holding time, which increases decoherence probability and affects the downstream key rate introducing statistical distributions and QBER variability. Quantum memory device modeling of this type is of benefit as it can be integrated into various protocol variations (e.g., standard MDI-QKD \cite{Abruzzo2014}, entanglement swapping \cite{Semenenko2022}  or different error-correction code rates \cite{Chelluri2026}). Stochastic models also track individual "events" which, in complex schemes, can highlight how coherence time limits performance under realistic timing mismatch. Analytical formulas for asynchronous cases can quickly become intractable or non-representative, which paves the way for stochastic modeling.

\begin{figure*}[!t]
    \centering
   \includegraphics[width=0.85\textwidth,height=0.42\textheight,keepaspectratio]{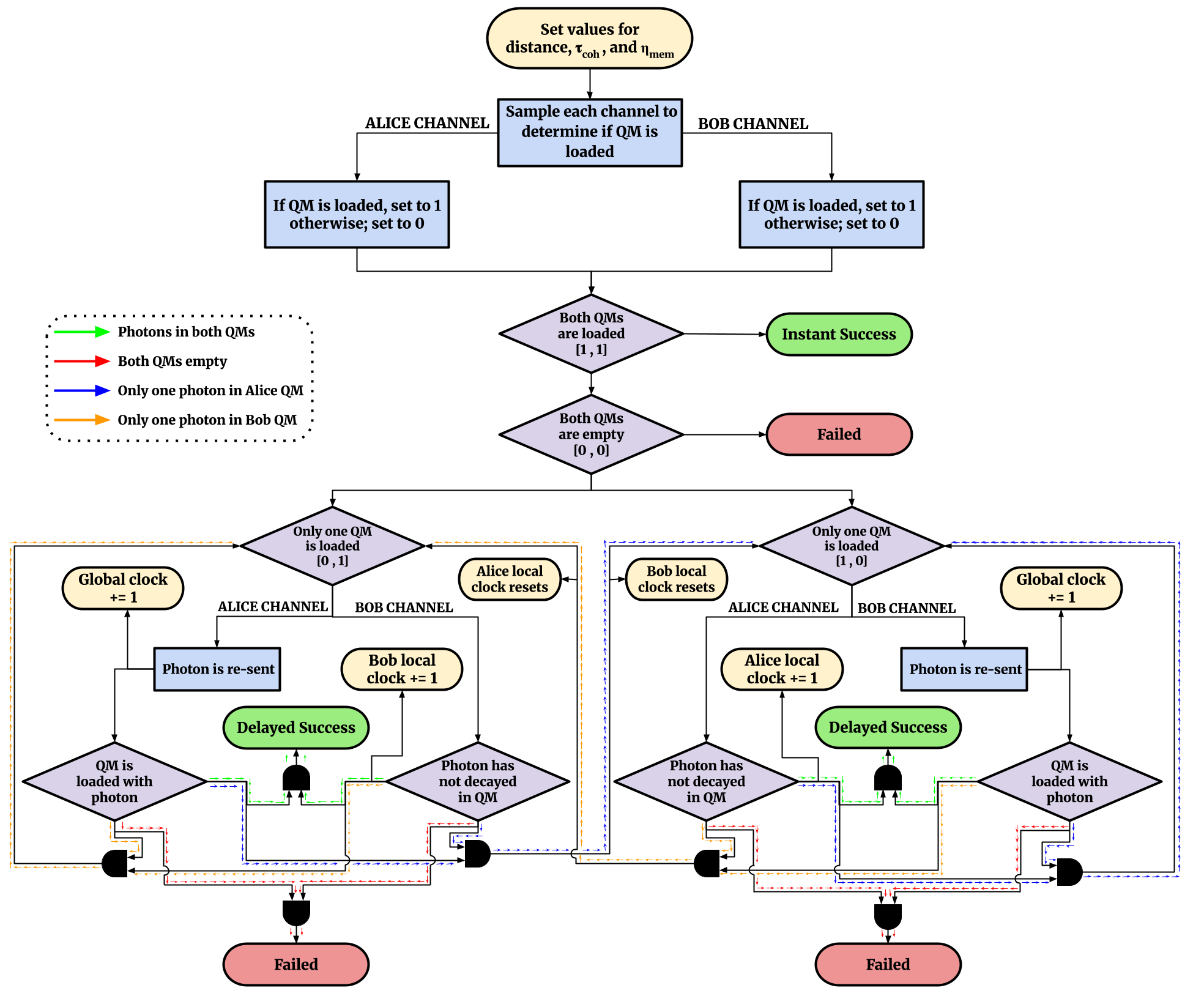}
   \caption{Logic flow for the stochastic modeling of MA-MDI-QKD key rates, utilizing the global/local clock scheme.}
   \label{fig:big_model}
\end{figure*}

To simulate MA-MDI-QKD key rates, our model accounts for the possibility of photonic states arriving either synchronously or asynchronously, where one QM can hold onto a photonic qubit while waiting for the second QM to be successfully loaded. The logic of our model is based on a global/local clock scheme depicted in Fig.~\ref{fig:big_model}, with Monte Carlo sampling undertaken for the memory loading and holding processes.

{The global and local clock framework was previously introduced in a memory-assisted entanglement distribution protocol \cite{Ahmed2026}. The present work extends that framework to MA-MDI-QKD. The new elements introduced here are heralded loading specific to MDI-QKD, SPS and WCP source models, an error-corrected throughput scaled by the mean global clock time and the global-clock latency statistics. In the synchronous limit where delayed successes are excluded, the stochastic model reduces to the deterministic gain expressions which are similar in structure to the analytical MA-MDI-QKD treatments of Panayi et al. \cite{Panayi2014} and Abruzzo et al. \cite{Abruzzo2014}. The agreement between the stochastic and deterministic synchronous results therefore serves as the limiting-case validation of the stochastic model.\par}

The stochastic modeling algorithm can be broken down as follows:
\vspace{0.1cm}
\begin{enumerate}
\item Each trial tracks photonic qubits from source emission through their individual channel, accounting for loss effects until arrival at Charlie. Monte Carlo sampling then determines whether each QM storage was successfully heralded.

\item One global clock unit is defined as the heralding round trip $t=L/v$. In the synchronous case, both QMs load successfully, the parties receive heralds and a BSM is performed after one time unit while both stored qubits are subjected to lifetime decay.

\item In the asynchronous case, a failed herald prompts a resend to one channel while the qubit in the second is subjected to decay with time. The core of the algorithm is that the local clock of each QM advances when it holds a qubit and then will reset when a reload is necessary. The global clock advances with each transmission attempt until both QMs hold a qubit to proceed to a delayed BSM or if the trial fails (both QMs empty).

\item The complete trial then resets. Here, full asynchronous results were obtained after the point-to-point variations between iterations had stabilized and further iterations produced no discernible change in the calculated error-corrected throughput rates. The fraction of successful BSM events is used to estimated the total gain $Q$, with each success retaining its recorded global clock time for a mean rate scaling calculation.

\end{enumerate}

\section{Modeling Considerations}  \label{Sec:4}

In developing MDI-QKD systems on a metropolitan scale, one may look to separate protocol optimization from security proofs such that hardware-related factors can be isolated as precursors before considering more complex effects. This can include optimizing asymmetric paths in quantum networks to test causes of latency before applying a decoy-state post-processing layer.  Described below are the major considerations treated in our model.

\subsection{Qubits based on weak coherent pulses}

In MDI-QKD schemes, the qubit sources held by Alice and Bob in Fig. \ref{CitySchem} are often one of two main types: a single photon source (SPS) or weak coherent pulses (WCP). An ideal single-photon source removes the multiphoton vulnerability of attenuated coherent states, making it robust against photon number splitting (PNS) attacks, although the security of a complete implementation still depends on the protocol assumptions and the remaining devices \cite{Bennett1984}. However, a practical and high-rate single-photon source remains a technological challenge. Furthermore, interfacing with a memory requires added requirements of photonic mode matching consisting of central frequency, bandwidth, polarization etc. \cite{Macrae2012}. Hence, many experimental MDI-QKD implementations often rely on WCP. \rev{Practical source effects, such as spectral correlations in heralded single-photon sources, further reduce the two-photon interference visibility in MDI-QKD \cite{Zhan2023,Chen2026}.} Our treatment provides a hardware-level systems analysis and protocol-agnostic comparison, isolating memory bottlenecks prior to cryptographic privacy amplification.  For future models, producing secret key rates will require consideration of the multi-photon component of WCPs, which opens a vulnerability to PNS attacks, where an eavesdropper can passively split off ``extra" photons from a multi-photon pulse without disturbing the remaining photons \cite{Lutkenhaus2002}. \rev{Consequently, the WCP results presented here should not be interpreted as secure key rates, as multiphoton components cannot be treated as secure contributions without a decoy-state analysis \cite{LoDecoyStates2005}.}

\subsection{Field-deployable quantum memory system}

The QMs are buffers at the central node in a MA-MDI-QKD system and serve to mitigate independent losses occurring in Alice and Bob's quantum channels.
 
Setting up central stations in a metropolitan environment will necessitate the availability of numerous, field-deployable QM devices \cite{Wang2022,Wang2025}. Ensemble-based QMs, such as warm vapor memories, can potentially fulfill these conditions \cite{Simon2011}. Warm vapor memories have now reported efficiencies up to 94\% \cite{Guo2025}, and, despite thermal motion, have shown coherence times of up to 1 ms \cite{Wang2022} along with fast repetition rates \cite{Finkelstein2018}. Furthermore, they have been shown capable of heralding in Raman schemes \cite{Michelberger2015} and are moving toward the same functionality in EIT systems \cite{Cheng2020}. Our model will assume the memories to be heralded with on-demand retrieval.  Loading will be probabilistic with storage functionality determined by the coherence decay time chosen in the model. 

Our model will assume that both Alice and Bob's QMs operate identically and store and release photonic qubits in the same spectral mode and temporal profile. In this demonstration of our model, we assume the errors in the protocol stem from the QMs. In other words, the polarization states of Alice and Bob's qubits have a finite probability of undergoing an error due to being held in the QM. \rev{Here, $E_{A/B}$ is taken as an unconditional error yield per trial for each memory, with $E_{A/B}=10^{-8}$. It represents memory-induced noise events, such as false heralds or readout noise, that occur independently of whether a signal photon was successfully stored and detected.} For stochastic analysis we will look at a range of QM efficiencies from $\eta_{mem}=0.1$, $0.5$ and $1$ and memory coherence times of $\tau_{\text{coh}}=0.1$, $0.5$ and $1$ ms. The QM coherence time will be modeled as an exponential degradation of the optical excitation inside the memory behaving as

\begin{equation}
    \eta_{coh}=e^{-t/\tau_{\text{coh}}}.
\end{equation}

\subsection{Propagation considerations}

The simulations here will focus on free space propagation channels and use an atmospheric loss value of $\alpha_{atm~loss}=0.1$ dB/km. For free space we will also include loss effects due to beam divergence.  A geometric loss term $\eta_{geo} (L) = \left( D/(2\theta L) \right)^2$
accounts for the reduction in received power as a function of the propagation distance $L$ which is dependent on the transmitting wavelength $\lambda$, beam waist $w_0$, beam divergence half-angle $\theta$ and receiver's aperture $D$. \rev{This expression is the ratio of the receiver aperture area to the cross-sectional area of the diverging beam at the receiver distance, where $\theta=\lambda/(\pi w_0)$ is the diffraction-limited half-angle divergence of a Gaussian beam with waist $w_0$ \cite{SalehTeich2007}, which is valid in the far field.} In our simulations, we chose values of  $\lambda=780$ nm, $w_0=3$ mm and receiver's aperture diameter $D=0.1$ m and optical collection $\eta_{coll}=0.7$ which is reachable using adaptive optics \cite{Scarfe2025}. The full channel transmission for a given length $L$ is then given by $\eta_t(L)=\eta_{geo}(L)\eta_{coll}10^{-
\alpha_{atm~loss}L/10}$.  These values were chosen to be in line with prior free space communication demonstrations \cite{Sit2017}. Throughout this work, $L$ denotes the full separation between Alice and Bob, such that each user-to-Charlie channel has length $L/2$ and the MA-MDI-QKD transmission per arm is $\eta_t(L/2)$, while the direct BB84 reference uses $\eta_t(L)$.

\subsection{Further technical limitations}

For systems utilizing polarized photons in the BSMs at the central station, active compensation or robust encoding schemes may be required to mitigate polarization distortions during transmission \cite{Tan2024}. Imperfect devices in a polarization-based BSM introduce deviations from ideal behavior including misalignment, mode mismatch, source fluctuations and beam splitter imperfections, and require careful treatment of BSM efficiency and error rates \cite{ZhongLo2015}.  Our implementation presented here will not consider BSM limitations. \rev{The numerical results therefore correspond to an idealized BSM benchmark. Imperfect two-photon interference between the photons retrieved from the two memories could be incorporated into the stochastic framework through the Hong-Ou-Mandel visibility $V$, which adds an optical error contribution of approximately $(1-V)/2$. The spectral, temporal and polarization distortion of a retrieved photon can depend on its storage time and $V$ can be sampled per event from the local clock of each memory. Polarization drift and its compensation in practical MDI-QKD \cite{Li2023} could also be treated in the same manner.}

Another major consideration is dark counts and stray light. Urban background from scattered sunlight \cite{Krzic2023} and artificial lighting at night \cite{Yastremski2025} raises the effective quantum bit error rate and motivates narrow-band filtering and time gating at the receiver. Acquisition and tracking systems \cite{Jaouni2025} can help maintain telescope alignment over multi-kilometer paths. Heralded QMs can suppress some false events, but they do not in general eliminate false heralding, memory readout noise, BSM detector dark counts, accidental coincidences or background photons. In the present implementation, these contributions are neglected for the MA-MDI-QKD system, and the results should be interpreted as an idealized hardware benchmark.

\section{Error-Corrected Throughput Analysis}\label{Sec:5}

To analyze the key rates, our model focuses on the domain of quantum hardware engineering and network performance analysis, specifically the memory-assisted physical layer throughput. Implementation of these rates will isolate performance as due to optical channel loss versus quantum bit errors. The error corrected rate focuses on analyzing the dynamics of the memory itself, such as how stochastic coherence decay $\tau_{coh}$ can impact synchronization independent of a specific privacy amplification algorithm or conservative worst-case security bounds. This is analogous to the hardware benchmark often specified in quantum repeater and memory-based architecture papers \cite{Semenenko2022}. We emphasize that the error-corrected throughput is distinct from a secure secret key rate. It accounts for the detected gain and the cost of error correction, but it does not include Eve's information, phase-error estimation, privacy amplification or decoy-state estimation.

\subsection{Rates using single photon states} 

When modeling assuming a SPS, the key rate will be calculated using Eq. \ref{Eq:1} consisting of three primary elements: total gain, error rate and the cost of error correction, expressed as follows \cite{Bebrov2024}:
\begin{equation}
R_{corr} = Q[1 - fh(QBER)]. \label{Eq:1}
\end{equation}
Where:
\begin{itemize}
    \item $R_{corr}$: Error-corrected throughput
    \item $Q$: Total Gain (probability of detecting a signal)
    \item $f$: error correction efficiency, taken to be 1 here
    \item $h(QBER)$: binary entropy function of QBER (Quantum Bit Error Rate) 
\end{itemize}
Here, $Q$ is the raw rate of signals that are recorded at the central station. The term $Q[1 - h(QBER)]$ describes the fraction of this raw rate that remains after accounting for errors and will be referred to as the error-corrected throughput, which has been utilized as a metric in prior hardware-based studies \cite{Mo2011}. Note that the binary entropy function is given by
$h(e) = -e \log_2(e) - (1 - e) \log_2(1 - e)$.

In MDI-QKD, the raw gain $Q$ can be calculated as follows:

\begin{equation}
Q_{SPS}=\frac{1}{2} \left( n \eta_{\text{t}} \eta_d \eta_{mem} e^{- \frac{t}{\tau_{\text{coh}}}} \right)^2. \label{Eq:2}
\end{equation}

\noindent Here, $n$ is the number of photons (set to 1 for SPS), $\eta_d$ is the detector efficiency and $\eta_{mem}$ is the total quantum memory efficiency. The term $\eta_t$ represents the free-space transmittance efficiency of each user-to-Charlie arm, $\eta_t(L/2)$, assuming that noise from background light sources is a negligible contribution. The modeling of the atmospheric inefficiency factors was discussed in the prior Section. 

Eq. \ref{Eq:2} calculates the contribution to the total gain for the realizations that yield two clicks when both QMs herald. The expression assumes both qubits sent by Alice and Bob undergo a BSM at Charlie's intermediate station in the presence of detector ($\eta_d$), QM ($\eta_{mem}$) and atmospheric transmittance ($\eta_t$) efficiencies. Since the QMs in our model are assumed to be heralded, we must consider only the realizations where both qubits are loaded in the QMs.  The $1/2$ factor preceding each term accounts for the maximum BSM efficiency of 50\%. 

In order to obtain the Quantum Bit Error Rate (QBER), we must first determine the rate of incorrect bits. \rev{In this idealized benchmark, detector dark counts are neglected (see Section~\ref{Sec:4}),} but the scheme remains vulnerable to errors generated by the memories themselves. We utilize a simplified static memory error \cite{Panayi2014} for both Alice's and Bob's memories, denoted as $e_{mem}$. \rev{Following the definition of $E_{A/B}$ in Section~\ref{Sec:4} as an unconditional error yield per trial, these erroneous events do not scale with the signal gain $Q$ and enter the QBER in the same manner as detector dark counts in the BB84 baseline of Section~\ref{Sec:6}. The probability that exactly one of the two memories introduces an error is}

\begin{equation} \label{Eq:3}
e_{mem}=2E_{A/B}(1-E_{{A/B}}),
\end{equation}

\noindent where $E_{A/B}$ is the unconditional error yield of each QM in the setup. The factor of 2 accounts for the fact there are two memories in the system. The QBER based on a stochastic gain is then calculated in our model as
\begin{equation} \label{Eq:4}
QBER=\frac{2E_{A/B}(1-E_{{A/B}})}{Q},
\end{equation}

\noindent \rev{which is the leading-order magnitude of $e_{mem}/(Q+e_{mem})$, valid for $e_{mem}\ll Q$. Hence, Eq. \ref{Eq:4} serves as a conservative estimate of the QBER. The error-corrected throughput values fall steeply and the approximation is adequate over the parameter range considered in the simulations. It can be seen, the QBER increases as the gain decreases with distance, and this behavior sets the threshold distances observed in the figures. The implementations of a conditional per-event memory error, which would instead contribute a distance-independent QBER floor, is discussed in Section~\ref{Sec:9}.}

\subsection{Rates using weak coherent pulses}
Many real-world implementations turn to WCP which are readily available from a laser source. Pulses composed of WCP have a Poissonian distribution of photon number states where the number of photons per pulse is distributed with probabilities to contain zero, one or more than one photon. By applying the Poisson distribution, the probability of photons in a WCP present in the QM at Charlie's station can be modeled as,

\begin{equation}\label{Eq:5}
P_{\mu}=1 - e^{-\mu \eta_d \eta_{mem} \eta_t}, 
\end{equation}

\noindent where $\mu$ is the mean photon number per pulse. Note that Eq. \ref{Eq:5} can be explained by unity minus the probability that a photon does not exist. Therefore, Eq. \ref{Eq:2} is modified as: 

\begin{equation} \label{Eq:6}
Q_{WCP} = \frac12 \left( 1 - e^{-\mu \eta_{\text{t}} \eta_d \eta_{mem} e^{-t/\tau_{\text{coh}}}} \right)^2.
\end{equation}

\noindent The above expression is now used with Eq. \ref{Eq:4} and Eq. \ref{Eq:1} to calculate the error-corrected throughput.  

For a scaled rate over transmission distance $L$, the attempt rate $\nu$ is determined by the propagation speed $v$ of the optical states in the medium so that $\nu=v/L$ and

\begin{equation}\label{Eq:7}
    R_{total}=\nu\cdot R_{corr}.
\end{equation}

{One global clock unit, $t=L/v$, corresponds to the heralding round trip $2(L/2)/v$ on each arm. This attempt rate describes a single-mode, stop-and-wait architecture and is therefore a conservative assumption based on the limits of propagation. A system operating with either a higher source clock or temporal multiplexing could increase the rate \cite{FanYuan2026}, by approximately the number of modes available. The absolute rates reported here can be rescaled accordingly, while the relative trends with memory efficiency, coherence time and waiting-time statistics would, in general, remain unchanged.\par}

\section{Key distribution rates with synchronous arrival times}\label{Sec:6}

In the synchronous scheme, each trial, a loading is attempted in each QM and the trial is discarded if either loading fails. The gain rate yields the shortest global clock time when time tagging successful BSM, as we do not wait for the other QM to load.  Once heralded, the central station notifies Alice and Bob, who return a message to trigger the BSM. Each QM therefore holds its qubit for one global clock count corresponding to the heralding round trip over its arm of length $L/2$, with $t=L/v$. For comparison, the lifetime decay $e^{-t/\tau_{coh}}$ is factored into both the deterministic key rate equations and the stochastic modeling. This scenario is closely related to a purely photonic MDI-QKD setup with a central station and will serve as an initial evaluation of our model.

We compare our model against a direct BB84 protocol whose rate is impacted by the presence of dark counts. In BB84 the gain is modeled as $Q_{BB84}=\eta_t(L)\eta_d$, where $\eta_t(L)$ corresponds to the transmission over the full distance $L$ separating Alice and Bob. The corresponding BB84-related QBER is $QBER_{BB84}=P_d/(Q_{BB84}+P_d)$ where $P_d$ is the dark count probability and is assumed to be $2.5\times10^{-4}$, which is typical of state-of-the-art photon counting modules at 780~nm. Dark counts are included in the BB84 baseline but largely neglected for the MA-MDI-QKD system and therefore the comparison should be regarded as a benchmark in the presence of idealized hardware.

\begin{figure}[!htbp]
    \centering
    \includegraphics[width=\columnwidth,height=0.26\textheight,keepaspectratio]{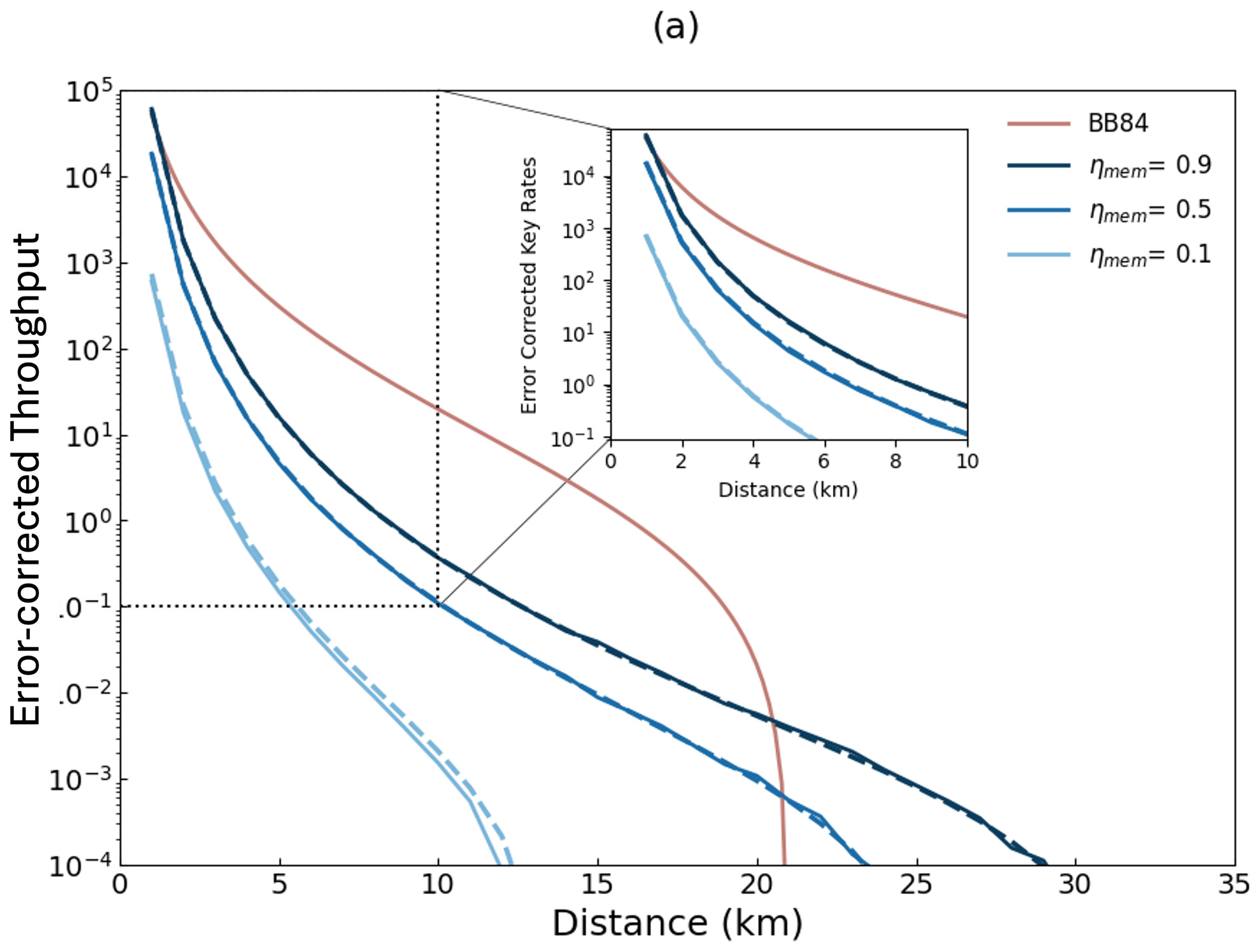}
    \includegraphics[width=\columnwidth,height=0.26\textheight,keepaspectratio]{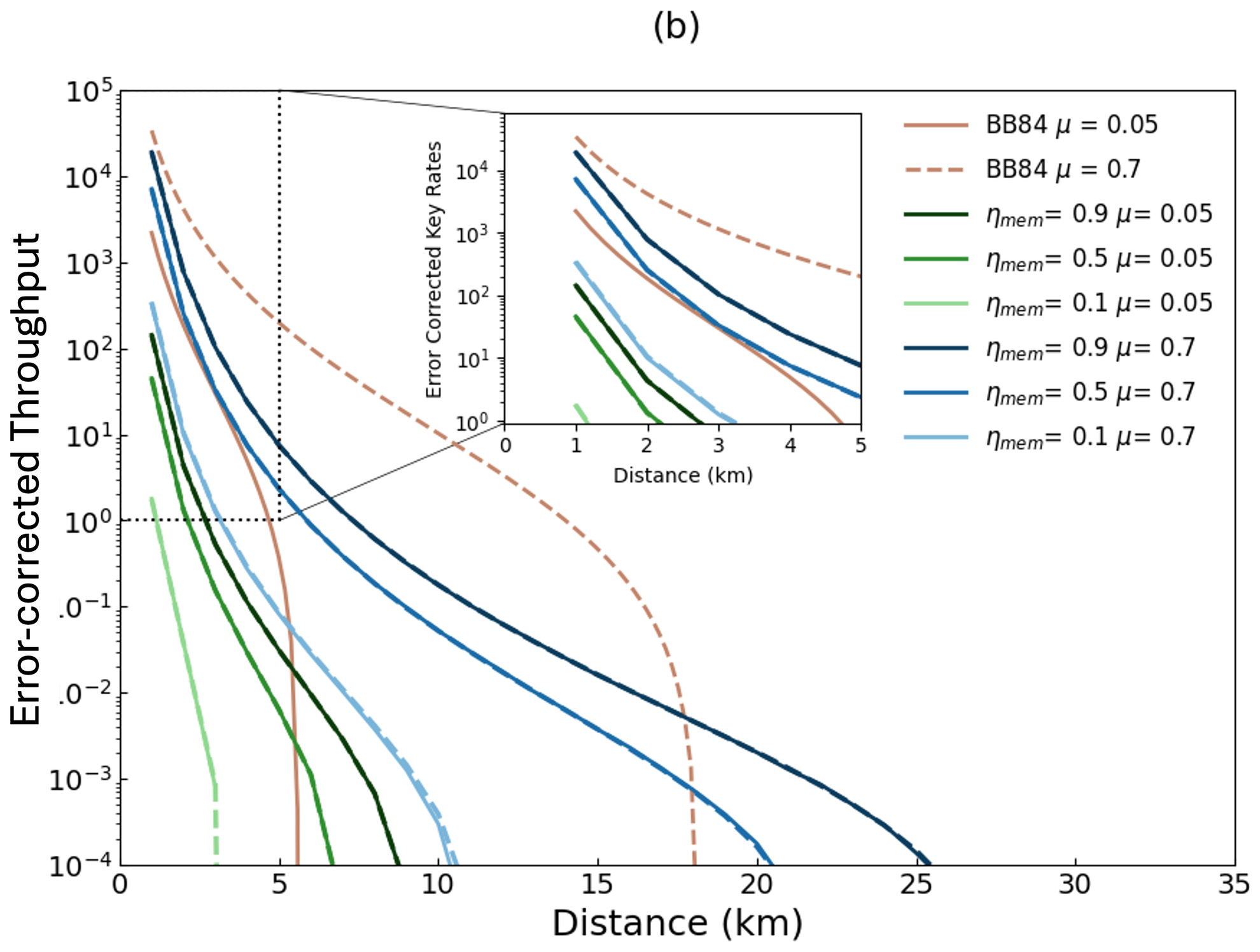}
    \caption{(a) Error-corrected throughput vs distance using SPS in a synchronous MA-MDI-QKD protocol with different QM efficiencies for deterministic (dashed) and stochastic (solid) based rates. Memory efficiencies of $\eta_{mem}$ of 0.9, 0.5, 0.1 are compared to direct BB84 (no memories). (b) Same as (a) but with WCP sources of $\mu=0.7$ and $\mu=0.05$.}
    \label{sync_graphs_1}
\end{figure}

\begin{figure}[!htbp]
    \centering
    \includegraphics[width=\columnwidth,height=0.26\textheight,keepaspectratio]{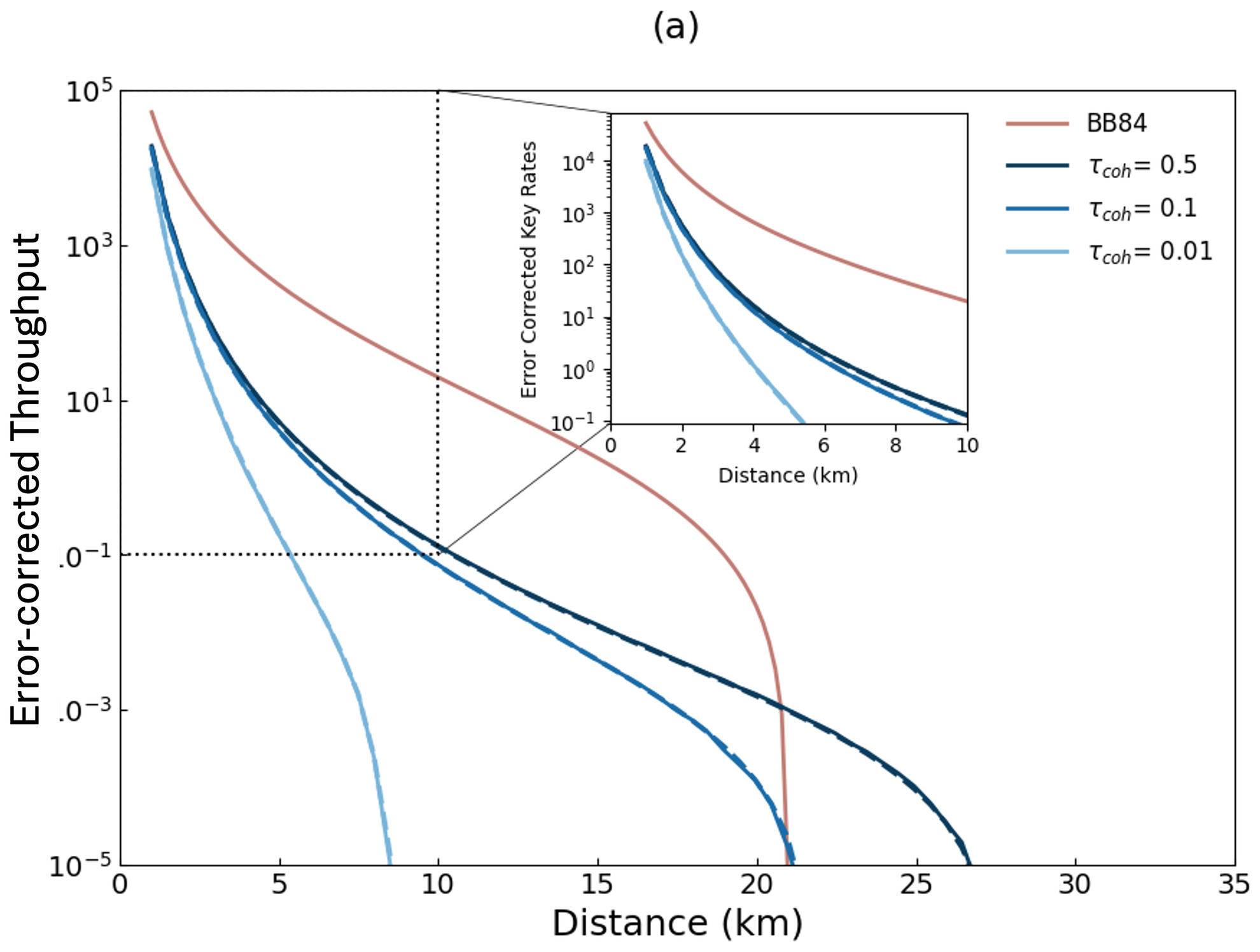}
    \includegraphics[width=\columnwidth,height=0.26\textheight,keepaspectratio]{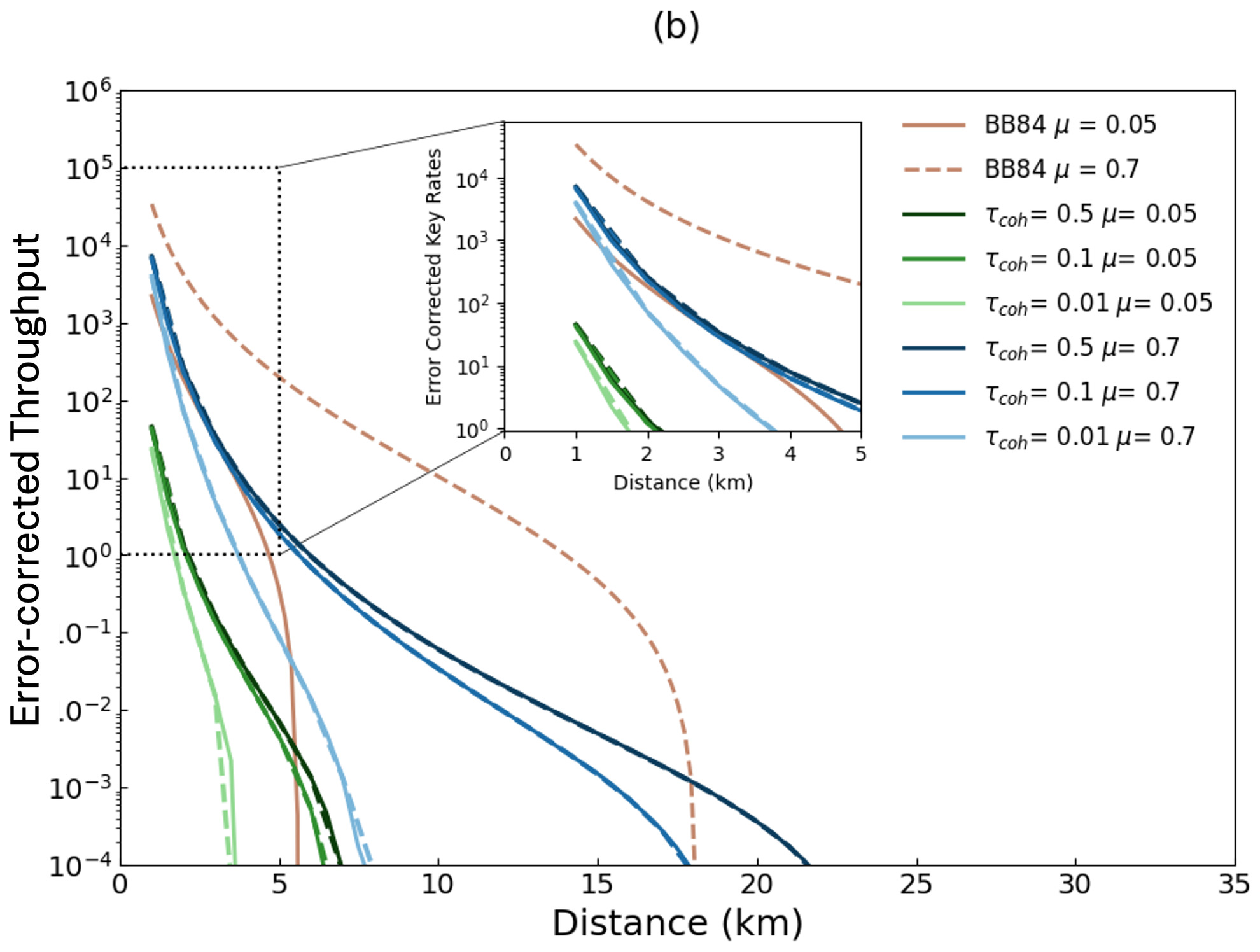}
    \caption{(a) Error-corrected throughput vs distance using SPS in a synchronous MA-MDI-QKD protocol with different QM efficiencies for deterministic (dashed) and stochastic (solid) based rates. Coherence times $\tau_{coh}$ of 0.5 ms, 0.1 ms, 0.01 ms are compared to direct BB84 (no memories). (b) Same as (a) but with WCP sources of $\mu=0.7$ and $\mu=0.05$.}
    \label{sync_graphs_2}
\end{figure}

The plots shown in Figs. \ref{sync_graphs_1} and \ref{sync_graphs_2} represent the error-corrected throughput, representative of hardware-related performance, and assume a scenario where no eavesdropper is present. The curves show the results from our stochastic model (solid) assuming no delayed successes compared to those using the deterministic expressions (dashed) based on Eqs. \ref{Eq:2} and \ref{Eq:6}. We model this synchronous scenario for variable QM efficiencies and a fixed coherence time of $\tau_{coh}=0.25$ ms which is shown in Fig. \ref{sync_graphs_1}. Similarly we also plot variable coherence times with a fixed QM efficiency of $\eta_{mem}=0.5$ as shown in Fig.~\ref{sync_graphs_2}. Rates are plotted for distances of up to 35 km, which can be typical of the span of a greater metropolitan area.
The Monte Carlo methods are validated due to the agreement between the stochastic and deterministic rates, serving as a starting point for extrapolations of the model where closed-form treatments become less tractable. \rev{This agreement provides the limiting-case correspondence with the analytical expressions introduced in Section~\ref{Sec:3}.} Note that the demand of a synchronous arrival time shown here will more often result in lower rates than a no-memory case due to the non-unity QM efficiency.

Figs. \ref{sync_graphs_1} and \ref{sync_graphs_2} highlight higher rates for SPS sources compared to WCP. This can be attributed to the probabilistic differences between the two light states. SPS contains singular distributions while a WCP source contains a non-zero probability of emitting zero photons. This translates to fractions of the transmitted pulses not registering successful counts, even if in principle the pulse could have reached the central station.  The synchronous rates surpass the BB84 protocol at around 21 km for SPS and 6 km and 18 km for WCP of mean photon numbers $\mu=0.05$ and $\mu=0.7$ respectively. When extracting the QBER versus distance the average QBER remains below the 11\% threshold for QM parameters of $\tau_{coh}=500~\mu$s and $\eta_{mem}=0.5$ for 20 km when using SPS and distances of 17 km for the cases of WCP of $\mu=0.7$. Note that the WCP curves represent error-corrected throughput and not secure key rates.

\section{Key distribution rates with asynchronous arrival times}\label{Sec:7}

We now turn our analysis to an asynchronous operation that considers delayed arrival times in addition to the synchronous ones. Alice and Bob send their qubits to the central station for loading into their specified QMs which are heralded. Upon Alice and Bob receiving notification of the herald from the QM, they can choose whether or not to resend their qubit.  The corresponding lifetime of the QM holding a qubit is stochastically modeled while waiting for the arrival of a qubit in the other channel.  Successes are recorded according to a BSM measurement after successful arrival of a qubit in both channels has been confirmed. For each success, the global clock time is recorded when both QMs have been loaded. The scheme fails when, at any time, neither Alice's nor Bob's QM holds a photon, and the algorithm shown in Fig.~\ref{fig:big_model} restarts.
The global/local clock scheme implemented here is directly applicable to entanglement distribution schemes \cite{Ahmed2026}.

The time tags of successful error-corrected detection events are tabulated using the global/local clock scheme detailed in Section \ref{Sec:3}. For a particular distance $L$ our algorithm will calculate the mean number of global clock units $\langle m \rangle$. With this modification  Eq. \ref{Eq:7} becomes

\begin{equation}\label{Eq:8}
    R_{total}=\frac{\nu}{\langle m \rangle}\cdot R_{corr}.
\end{equation}

\begin{figure}[!htbp]
    \centering
    \includegraphics[width=\columnwidth,height=0.26\textheight,keepaspectratio]{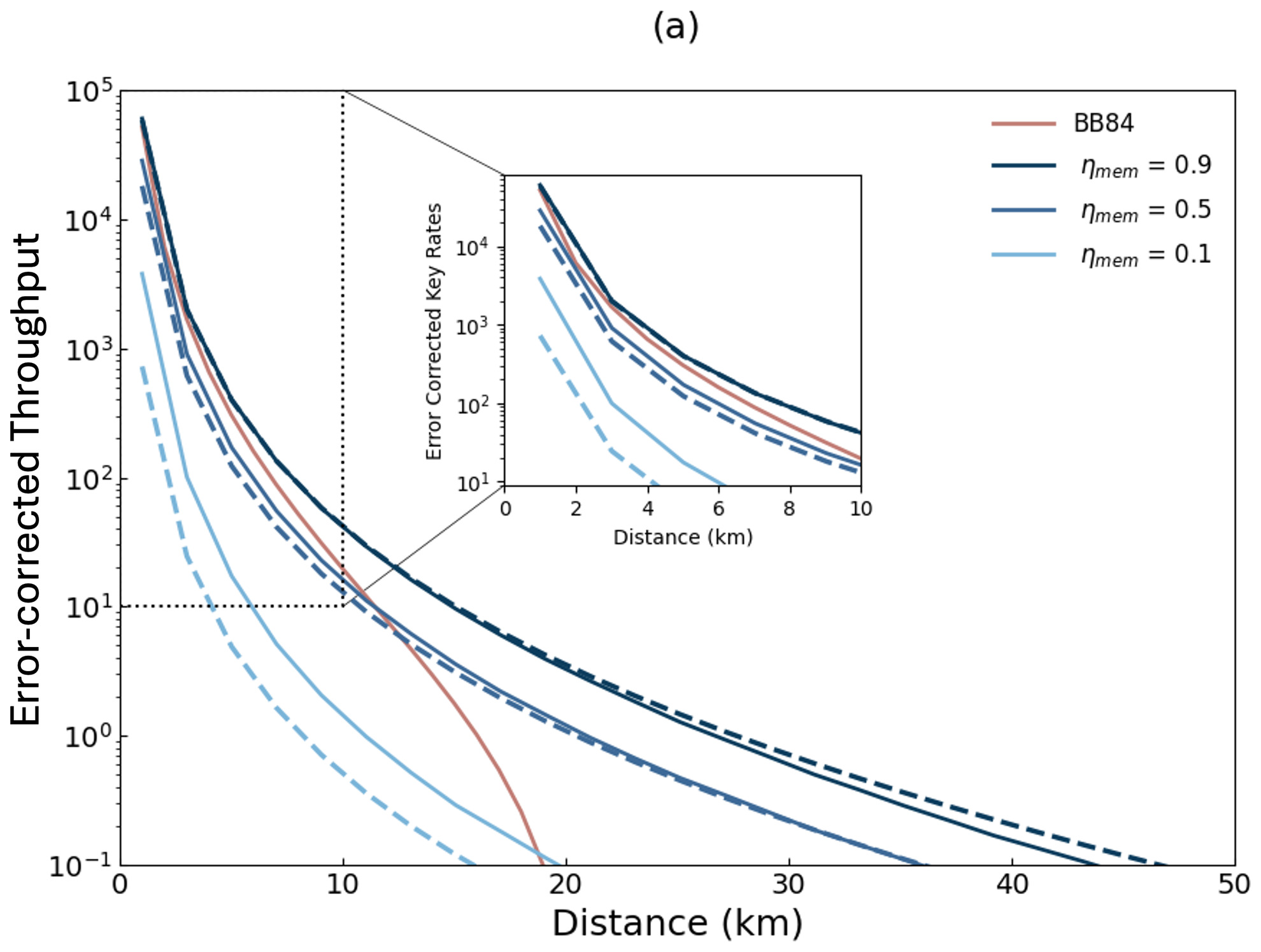}
    \includegraphics[width=\columnwidth,height=0.26\textheight,keepaspectratio]{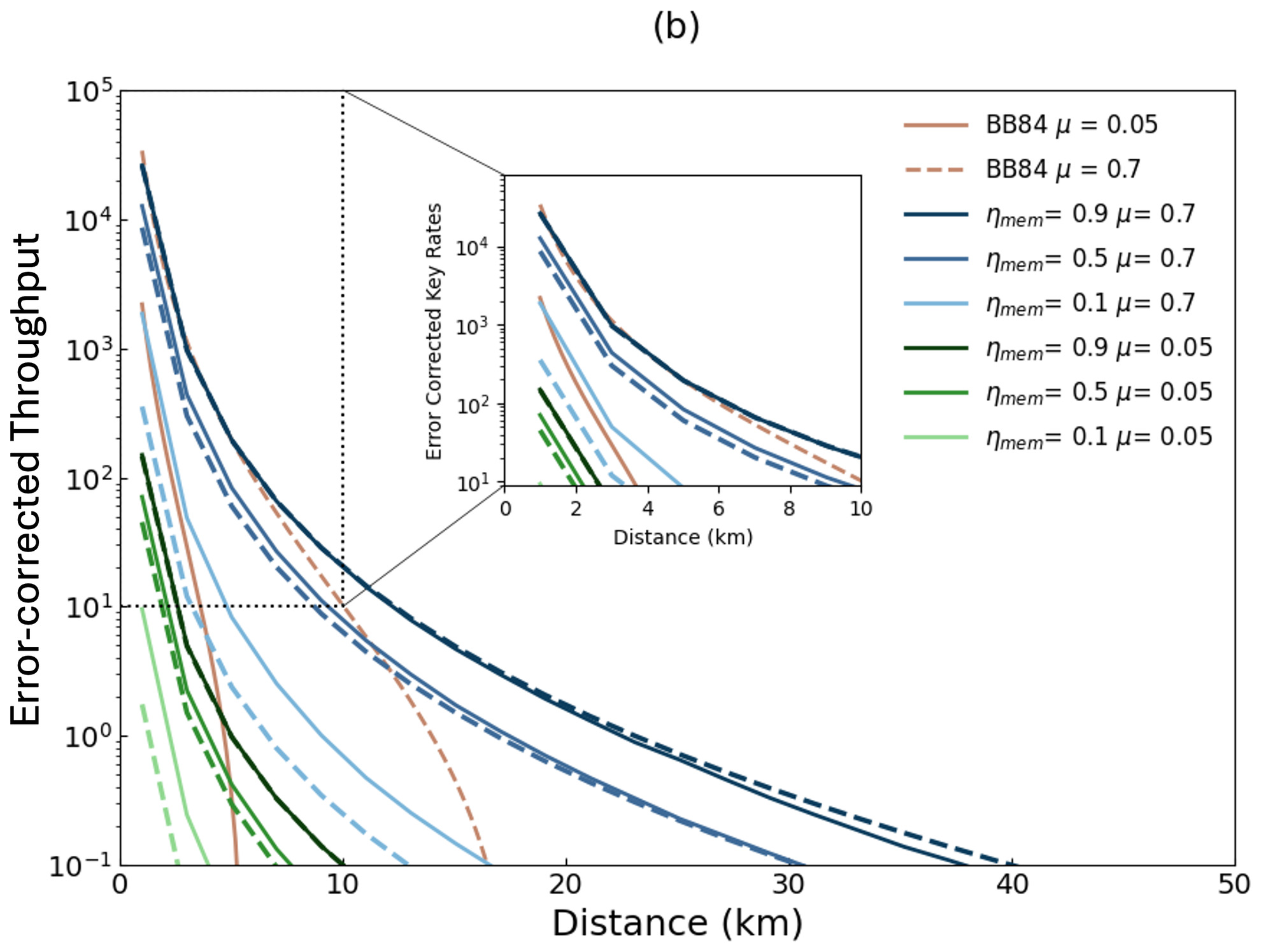}
    \caption{Deterministic \& stochastic MA-MDI-QKD vs BB84 modelling. (a) Error-corrected throughput vs distance using ideal SPS in an asynchronous MA-MDI-QKD protocol with different QM efficiencies. The deterministic results (dashed) correspond to a simplified model for synchronous cases with a $e^{-\alpha L/2}$ dependence for comparison against the stochastic results (solid). Curves represent QM efficiencies $\eta_{mem}$ of 0.9, 0.5, 0.1 with coherence time $\tau_{coh}$ kept constant at 0.25 ms and compared to direct BB84 (no memories). (b) Same as (a) but with WCP sources of $\mu=0.7$ and $\mu=0.05$.}
    \label{async_graphs_1}
\end{figure}

\begin{figure}[!htbp]
    \centering
    \includegraphics[width=\columnwidth,height=0.26\textheight,keepaspectratio]{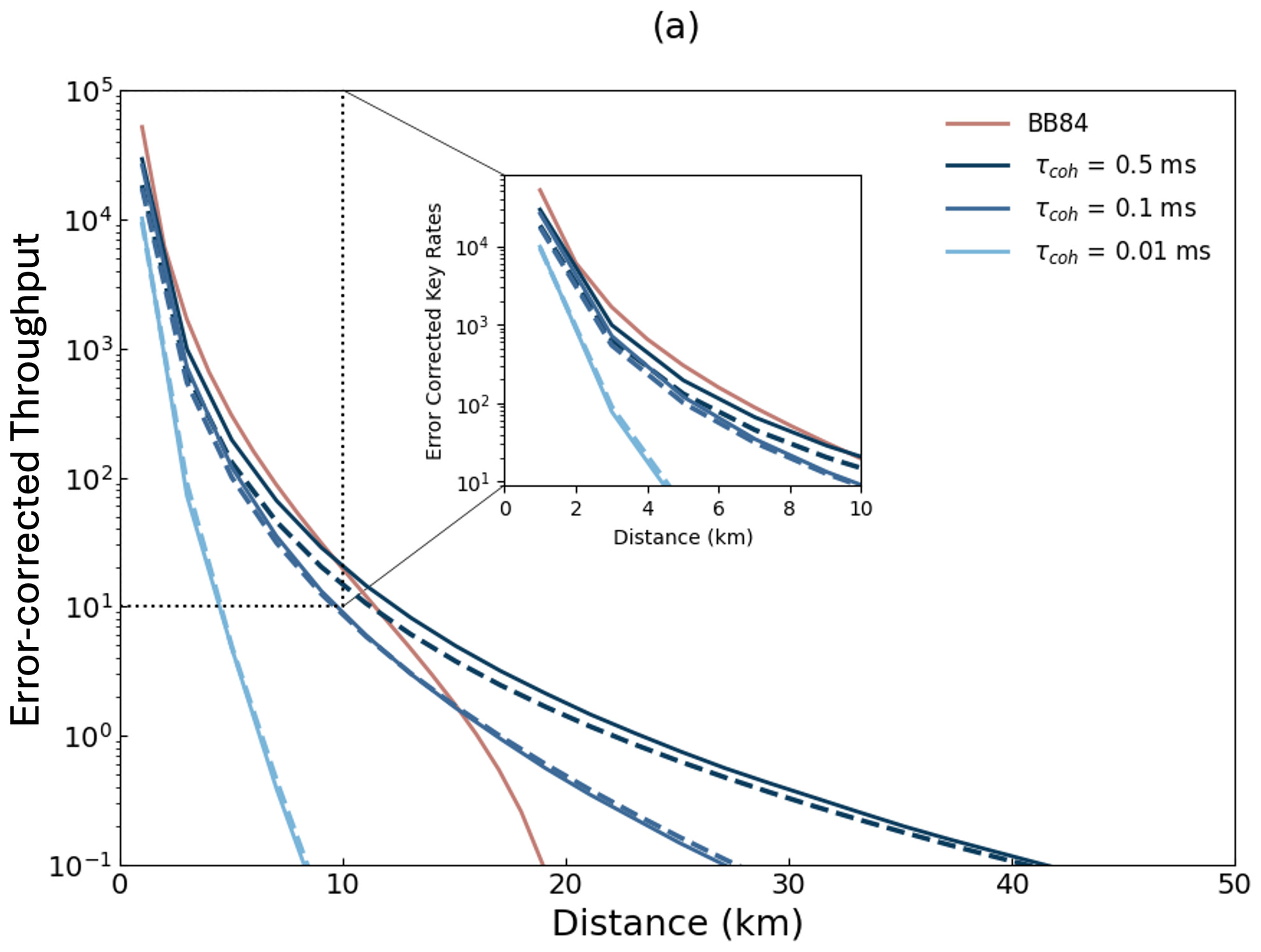}
    \includegraphics[width=\columnwidth,height=0.26\textheight,keepaspectratio]{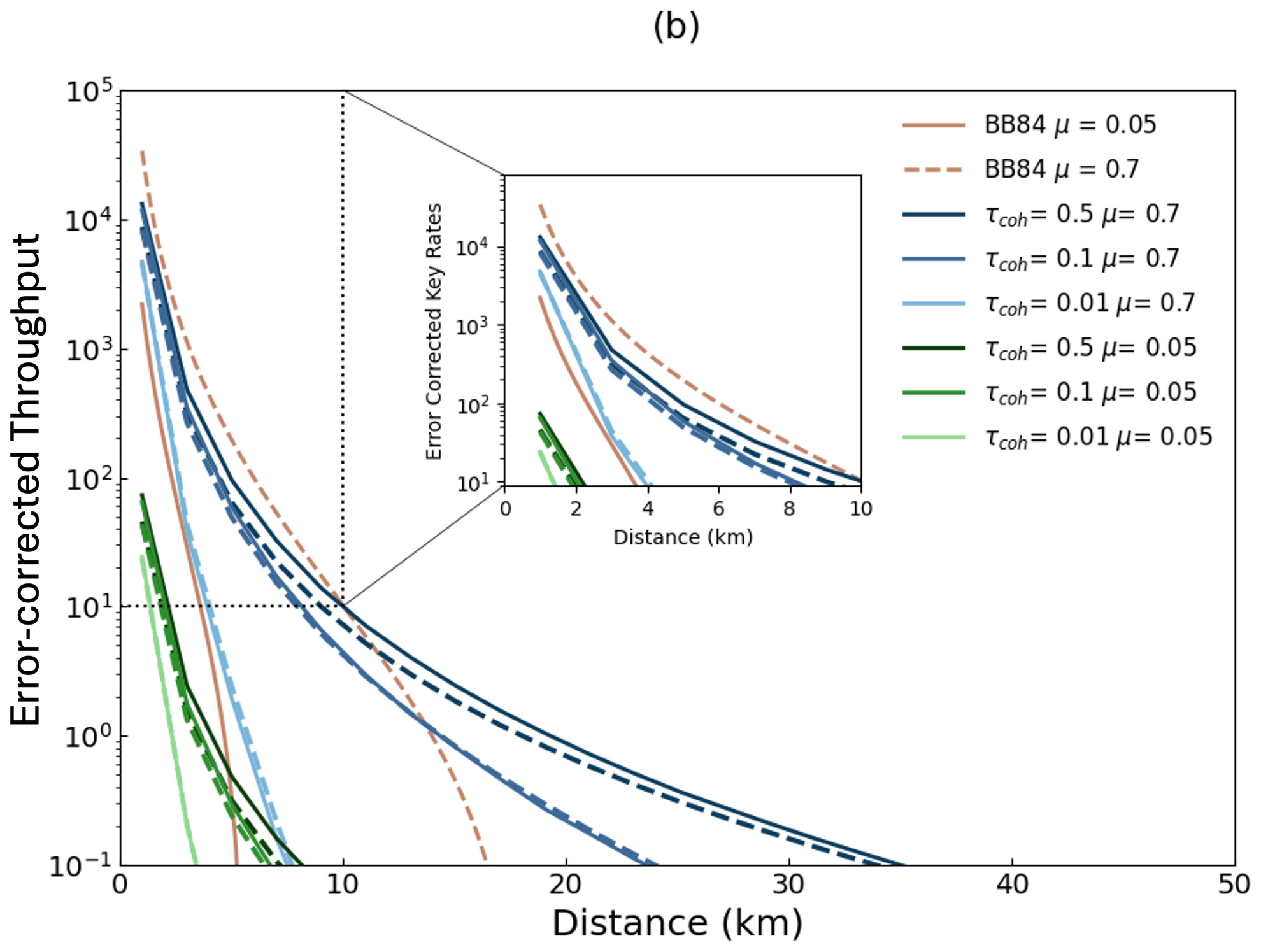}
      \caption{Deterministic \& stochastic MA-MDI-QKD vs BB84 modelling. (a) Error-corrected throughput vs distance using ideal SPS in an asynchronous MA-MDI-QKD protocol with different QM efficiencies. The deterministic results (dashed) correspond to a simplified model for synchronous cases with a $e^{-\alpha L/2}$ dependence and compared to the stochastic results (solid). Curves represent coherence times $\tau_{coh}$ of 0.5 ms, 0.1 ms, 0.01 ms with QM efficiency $\eta_{mem}$ kept constant at 0.5 and compared to direct BB84 (no memories). (b) Same as (a) but with WCP sources  of $\mu=0.7$ and $\mu=0.05$.}
    \label{async_graphs_2}
\end{figure}

\noindent Figs.~\ref{async_graphs_1} and \ref{async_graphs_2} highlight the results of the stochastic modeling where stochastic rates are plotted against a curve showing $Q$ and $QBER$  with an idealized $e^{-\alpha L/2}$ as an approximated dependence of the deterministic behavior. Fig.~\ref{async_graphs_1} shows the error-corrected throughput for a fixed QM coherence time of $\tau_{coh}=0.25$~ms and varying QM efficiencies, and Fig.~\ref{async_graphs_2} for a fixed QM efficiency $\eta_{mem}=0.5$ and variable coherence times with rates shown for distances up to 50 km.

In an asynchronous setup, the mean global clock time required for a successful BSM extends due to the possibility of resending qubits. An interesting feature is that the stochastic model outperforms the model that only counts synchronous cases but with a $e^{-\alpha L/2}$ distance dependence at low memory efficiencies for both SPS and WCP. This can be attributed to the lower demands on asynchronous arrival versus synchronous. This outperformance then decreases with higher memory efficiencies due to the averaging over the global clock time that consequently occurs at longer communication distances. Extracting the QBER values shows it remains below the 11\% threshold for QM parameters of $\tau_{coh}=100~\mu$s and $\eta_{mem}=0.5$ out to 59~km when using SPS and 52~km for WCP with $\mu=0.7$.

As can be observed in Figs.~\ref{sync_graphs_1}--\ref{async_graphs_2}, the curves identify a crossover distance beyond which memory assistance becomes advantageous. At short range, direct BB84 without memories yields higher rates, as shown in the insets of Figs.~\ref{sync_graphs_1} and \ref{sync_graphs_2}. At longer range, MA-MDI-QKD overtakes BB84 at near 21~km for SPS in the synchronous case and near 15--18~km for WCP with $\mu=0.7$ and at still shorter distances for high-efficiency asynchronous operation. Note that the error models used in these plots are asymmetric as errors in BB84 arise from dark counts, whereas errors in MA-MDI-QKD arise from an unconditional memory error yield. Therefore, the quoted crossovers represent an idealized hardware benchmark showing where a memory-assisted scheme can be advantageous under the parameters used here, rather than a direct comparison between deployable systems.  With the idealized noise assumptions for the MA-MDI-QKD system, the crossover distances are representations of the best-case estimated rates.

The asynchronous error-corrected throughput degrades at a reduced slope relative to the synchronous scheme. \rev{This can be attributed to the independent loading of each arm of length $L/2$. The synchronous scheme requires simultaneous success over both arms, so its gain scales as $\eta_t(L/2)^2\approx\eta_t(L)$, whereas the asynchronous scheme allows each arm to succeed in separate attempts, giving the $\sim\sqrt{\eta_t(L)}$ scaling of the transmission probability.} Comparison between the synchronous and asynchronous scenarios in Figs.~\ref{sync_graphs_1} vs.~\ref{async_graphs_1} and Figs.~\ref{sync_graphs_2} vs.~\ref{async_graphs_2} directly shows this impact where synchronous rates typically fall below practical thresholds by about 20--30~km, whereas asynchronous rates remain usable out to roughly 40--50~km. At a representative metropolitan distance of 25~km, the asynchronous case yields an enhancement of order $10^{3}$ relative to the synchronous scheme.

\begin{figure}[!htbp]
    \centering
    \includegraphics[width=\columnwidth,height=0.28\textheight,keepaspectratio]{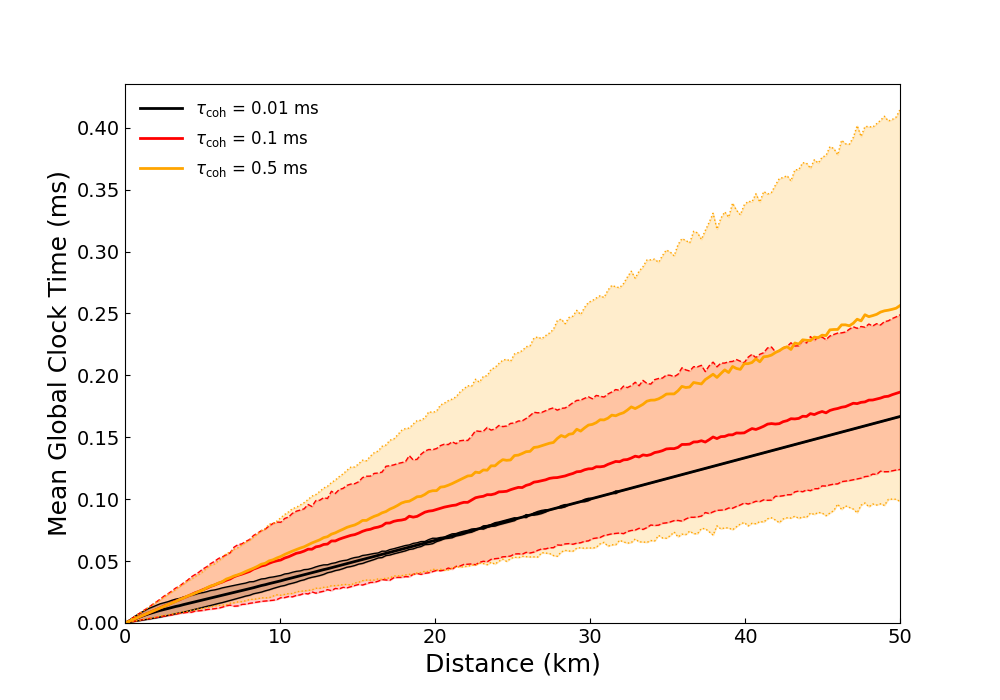}
    \caption{Mean global clock times vs distance for
    coherence times $\tau_{coh}$ of 0.01 ms, 0.1 ms, 0.5 ms. The shaded area represents the spread of each plot.}
    \label{mean_gc}
\end{figure}

Finally we highlight a metric often missing from key-rate-only analyses. Fig.~\ref{mean_gc} shows the mean global clock time and its standard deviation versus distance for each QM coherence time. The standard deviation represents the event-to-event latency variability and does not represent the statistical uncertainty in the mean. The mean wait time grows approximately linearly with distance, while the variance increases with longer $\tau_{\mathrm{coh}}$. For $\tau_{\mathrm{coh}}=0.01$~ms the spread collapses beyond about 20~km, since short coherence times leave only instant successes at long range. For $\tau_{\mathrm{coh}}=0.5$~ms the longer hold times yield standard deviations of roughly 0.08--0.40~ms, highlighting the implications for latency variability.

The stochastic approach exemplified here can be extended to MA-MDI-QKD topologies with multiple channels connected to a central hub that removes the need for dedicated channels between every party. \rev{Such architectures have been demonstrated experimentally for MDI-QKD networks in which multiple users communicate through an untrusted measurement node \cite{FanYuan2022}. Our framework would be of high value in models of networks that include node-dependent sources, channels and memory parameters as oppose to Alice and Bob holding identical subsystems.}

{More generally, future quantum networks are unlikely to consist of identical nodes when a protocol is realized \cite{Lu2025}. The global and local clock framework will allow a natural generalization to such heterogeneous networks where each node can be assigned its own local clock, memory efficiency, coherence time, channel loss, source rate and processing delay, while the global clock will be able to coordinate BSM scheduling at the hub. Since the model is event driven, no closed-form symmetry would be required between users.\par}

\section{Future model considerations}\label{Sec:9}

Future versions of this tool could look to incorporate multiple, more advanced considerations. This includes noise models that encompass additional sources of stray light from background sunlight and urban illumination.  Photons from these unwanted sources affect the MDI-QKD rate and have a direct impact on the achievable key rate \cite{Krzic2023}.

Another inclusion is the detailed imperfections of the BSM apparatus at the central station. Specifically, the propagation of waveplate uncertainties and beam splitter limitations on the BSM correlations \cite{ZhongLo2015}. This is critical as bit errors caused by an eavesdropper, Eve, are practically indistinguishable from those caused by imperfect equipment. The memory error model could also be extended to include a conditional per-event error probability, which would add a distance-independent contribution to the QBER alongside the unconditional error yield that was implemented here.

Future model iterations could involve channel specific effects. For example in free space effects this would include atmospheric turbulence that introduces random fluctuations in the refractive index causing $\vec{k}$ vector movement, beam divergence and scattering.  These effects can severely degrade the quality of the received quantum states and have been a topic of much investigation on metropolitan scales \cite{Scarfe2025}. 

The work presented here assumed distribution rates using SPS and then also WCP as a rate comparison in the case of SPS being unavailable. However, when WCPs are employed, a true determination  of the secret key rate requires the use of decoy states \cite{LoDecoyStates2005} to safeguard against potential photon number splitting attacks. This will be investigated in future work.

This preliminary demonstration of our simulation algorithm shows its value as a tool for predicting the performance of MA-MDI-QKD network topologies. This model can be extended as the primary function of MA-MDI-QKD schemes where an intermediate node serves as a hub for numerous solitary parties. 

\section{Conclusion}\label{Sec:10}

In this paper, we have \rev{applied the global/local clock framework \cite{Ahmed2026} to provide} a simplified model for error-corrected throughput in a MA-MDI-QKD system based on stochastic processes and a global/local clock timing system for each QM.  This permits rate calculation for asynchronous state arrivals and the tabulation of delayed successes.  With simulating metropolitan distances in mind, we have shown how this model provides insights into system performance under parameters including QM efficiency $\eta_{mem}$ and coherence time $\tau_{coh}$ to serve as a predictive tool in practical applications. The global-clock latency statistics allow hardware and network designers to trade memory efficiency, coherence time and wait-time variance for metropolitan distance MA-MDI-QKD. The reported rates correspond to an idealized hardware benchmark of error-corrected throughput rather than secure secret key rates.

The model presented sets a foundation based on stochastic modeling of quantum memories and enables future model iterations that add computational complexity, including dependent factors and asymmetric channels to the central MDI-QKD station.  The focus on error-corrected throughput moves toward optimization based on hardware-related factors that can be isolated before applying decoy-state post-processing for security proofs. This can be instrumental in advancing the design, deployment and overall understanding of future MA-MDI-QKD networks. Predictive modeling tools such as this will become increasingly relevant for global quantum networks composed of fibre, free space and satellite links \cite{Yastremski2025}; hence our model serves as a powerful complement to existing quantum network simulators.

\begin{acknowledgments}
The authors acknowledge the NSERC USRA and NSERC Discovery program for funding support.
\end{acknowledgments}

\bibliography{PhysicsLettersASubmission,MA-MDI-QKD_Revision_Additional_References}

\begin{thebibliography}{58}%
\makeatletter
\providecommand \@ifxundefined [1]{%
 \@ifx{#1\undefined}
}%
\providecommand \@ifnum [1]{%
 \ifnum #1\expandafter \@firstoftwo
 \else \expandafter \@secondoftwo
 \fi
}%
\providecommand \@ifx [1]{%
 \ifx #1\expandafter \@firstoftwo
 \else \expandafter \@secondoftwo
 \fi
}%
\providecommand \natexlab [1]{#1}%
\providecommand \enquote  [1]{``#1''}%
\providecommand \bibnamefont  [1]{#1}%
\providecommand \bibfnamefont [1]{#1}%
\providecommand \citenamefont [1]{#1}%
\providecommand \href@noop [0]{\@secondoftwo}%
\providecommand \href [0]{\begingroup \@sanitize@url \@href}%
\providecommand \@href[1]{\@@startlink{#1}\@@href}%
\providecommand \@@href[1]{\endgroup#1\@@endlink}%
\providecommand \@sanitize@url [0]{\catcode `\\12\catcode `\$12\catcode
  `\&12\catcode `\#12\catcode `\^12\catcode `\_12\catcode `\%12\relax}%
\providecommand \@@startlink[1]{}%
\providecommand \@@endlink[0]{}%
\providecommand \url  [0]{\begingroup\@sanitize@url \@url }%
\providecommand \@url [1]{\endgroup\@href {#1}{\urlprefix }}%
\providecommand \urlprefix  [0]{URL }%
\providecommand \Eprint [0]{\href }%
\providecommand \doibase [0]{https://doi.org/}%
\providecommand \selectlanguage [0]{\@gobble}%
\providecommand \bibinfo  [0]{\@secondoftwo}%
\providecommand \bibfield  [0]{\@secondoftwo}%
\providecommand \translation [1]{[#1]}%
\providecommand \BibitemOpen [0]{}%
\providecommand \bibitemStop [0]{}%
\providecommand \bibitemNoStop [0]{.\EOS\space}%
\providecommand \EOS [0]{\spacefactor3000\relax}%
\providecommand \BibitemShut  [1]{\csname bibitem#1\endcsname}%
\let\auto@bib@innerbib\@empty
\bibitem [{\citenamefont {Bennett}\ and\ \citenamefont
  {Brassard}(1984)}]{Bennett1984}%
  \BibitemOpen
  \bibfield  {author} {\bibinfo {author} {\bibfnamefont {C.~H.}\ \bibnamefont
  {Bennett}}\ and\ \bibinfo {author} {\bibfnamefont {G.}~\bibnamefont
  {Brassard}},\ }\bibfield  {title} {\bibinfo {title} {Quantum cryptography:
  Public key distribution and coin tossing},\ }in\ \href@noop {} {\emph
  {\bibinfo {booktitle} {Proceedings of the IEEE International Conference on
  Computers, Systems and Signal Processing}}}\ (\bibinfo {address} {Bangalore,
  India},\ \bibinfo {year} {1984})\ pp.\ \bibinfo {pages}
  {175--179}\BibitemShut {NoStop}%
\bibitem [{\citenamefont {Xu}\ \emph {et~al.}(2020)\citenamefont {Xu},
  \citenamefont {Ma}, \citenamefont {Zhang}, \citenamefont {Lo},\ and\
  \citenamefont {Pan}}]{PracticalQKD2020}%
  \BibitemOpen
  \bibfield  {author} {\bibinfo {author} {\bibfnamefont {F.}~\bibnamefont
  {Xu}}, \bibinfo {author} {\bibfnamefont {X.}~\bibnamefont {Ma}}, \bibinfo
  {author} {\bibfnamefont {Q.}~\bibnamefont {Zhang}}, \bibinfo {author}
  {\bibfnamefont {H.-K.}\ \bibnamefont {Lo}},\ and\ \bibinfo {author}
  {\bibfnamefont {J.-W.}\ \bibnamefont {Pan}},\ }\bibfield  {title} {\bibinfo
  {title} {Secure quantum key distribution with realistic devices},\ }\href
  {https://doi.org/10.1103/RevModPhys.92.025002} {\bibfield  {journal}
  {\bibinfo  {journal} {Rev. Mod. Phys.}\ }\textbf {\bibinfo {volume} {92}},\
  \bibinfo {pages} {025002} (\bibinfo {year} {2020})}\BibitemShut {NoStop}%
\bibitem [{\citenamefont {Lo}\ \emph {et~al.}(2012)\citenamefont {Lo},
  \citenamefont {Curty},\ and\ \citenamefont {Qi}}]{LoMDIQKD2012}%
  \BibitemOpen
  \bibfield  {author} {\bibinfo {author} {\bibfnamefont {H.-K.}\ \bibnamefont
  {Lo}}, \bibinfo {author} {\bibfnamefont {M.}~\bibnamefont {Curty}},\ and\
  \bibinfo {author} {\bibfnamefont {B.}~\bibnamefont {Qi}},\ }\bibfield
  {title} {\bibinfo {title} {Measurement-device-independent quantum key
  distribution},\ }\bibfield  {journal} {\bibinfo  {journal} {Physical Review
  Letters}\ }\textbf {\bibinfo {volume} {108}},\ \href
  {https://doi.org/10.1103/physrevlett.108.130503}
  {10.1103/physrevlett.108.130503} (\bibinfo {year} {2012})\BibitemShut
  {NoStop}%
\bibitem [{\citenamefont {Valivarthi}\ \emph {et~al.}(2015)\citenamefont
  {Valivarthi}, \citenamefont {Lucio-Martinez}, \citenamefont {Chan},
  \citenamefont {Rubenok}, \citenamefont {John}, \citenamefont {Korchinski},
  \citenamefont {Duffin}, \citenamefont {Marsili}, \citenamefont {Verma},
  \citenamefont {Shaw}, \citenamefont {Stern}, \citenamefont {Nam},
  \citenamefont {Oblak}, \citenamefont {Zhou}, \citenamefont {Slater},\ and\
  \citenamefont {Tittel}}]{Valivarthi2015}%
  \BibitemOpen
  \bibfield  {author} {\bibinfo {author} {\bibfnamefont {R.}~\bibnamefont
  {Valivarthi}}, \bibinfo {author} {\bibfnamefont {I.}~\bibnamefont
  {Lucio-Martinez}}, \bibinfo {author} {\bibfnamefont {P.}~\bibnamefont
  {Chan}}, \bibinfo {author} {\bibfnamefont {A.}~\bibnamefont {Rubenok}},
  \bibinfo {author} {\bibfnamefont {C.}~\bibnamefont {John}}, \bibinfo {author}
  {\bibfnamefont {D.}~\bibnamefont {Korchinski}}, \bibinfo {author}
  {\bibfnamefont {C.}~\bibnamefont {Duffin}}, \bibinfo {author} {\bibfnamefont
  {F.}~\bibnamefont {Marsili}}, \bibinfo {author} {\bibfnamefont
  {V.}~\bibnamefont {Verma}}, \bibinfo {author} {\bibfnamefont {M.~D.}\
  \bibnamefont {Shaw}}, \bibinfo {author} {\bibfnamefont {J.~A.}\ \bibnamefont
  {Stern}}, \bibinfo {author} {\bibfnamefont {S.~W.}\ \bibnamefont {Nam}},
  \bibinfo {author} {\bibfnamefont {D.}~\bibnamefont {Oblak}}, \bibinfo
  {author} {\bibfnamefont {Q.}~\bibnamefont {Zhou}}, \bibinfo {author}
  {\bibfnamefont {J.~A.}\ \bibnamefont {Slater}},\ and\ \bibinfo {author}
  {\bibfnamefont {W.}~\bibnamefont {Tittel}},\ }\bibfield  {title} {\bibinfo
  {title} {Measurement-device-independent quantum key distribution: from idea
  towards application},\ }\href {https://doi.org/10.1080/09500340.2015.1021725}
  {\bibfield  {journal} {\bibinfo  {journal} {Journal of Modern Optics}\
  }\textbf {\bibinfo {volume} {62}},\ \bibinfo {pages} {1141} (\bibinfo {year}
  {2015})}\BibitemShut {NoStop}%
\bibitem [{\citenamefont {Ekert}(1991)}]{E91}%
  \BibitemOpen
  \bibfield  {author} {\bibinfo {author} {\bibfnamefont {A.~K.}\ \bibnamefont
  {Ekert}},\ }\bibfield  {title} {\bibinfo {title} {Quantum cryptography based
  on bell's theorem},\ }\href {https://doi.org/10.1103/PhysRevLett.67.661}
  {\bibfield  {journal} {\bibinfo  {journal} {Phys. Rev. Lett.}\ }\textbf
  {\bibinfo {volume} {67}},\ \bibinfo {pages} {661} (\bibinfo {year}
  {1991})}\BibitemShut {NoStop}%
\bibitem [{\citenamefont {Rubenok}\ \emph {et~al.}(2013)\citenamefont
  {Rubenok}, \citenamefont {Slater}, \citenamefont {Chan}, \citenamefont
  {Lucio-Martinez},\ and\ \citenamefont {Tittel}}]{Rubenok2013}%
  \BibitemOpen
  \bibfield  {author} {\bibinfo {author} {\bibfnamefont {A.}~\bibnamefont
  {Rubenok}}, \bibinfo {author} {\bibfnamefont {J.~A.}\ \bibnamefont {Slater}},
  \bibinfo {author} {\bibfnamefont {P.}~\bibnamefont {Chan}}, \bibinfo {author}
  {\bibfnamefont {I.}~\bibnamefont {Lucio-Martinez}},\ and\ \bibinfo {author}
  {\bibfnamefont {W.}~\bibnamefont {Tittel}},\ }\bibfield  {title} {\bibinfo
  {title} {Real-world two-photon interference and proof-of-principle quantum
  key distribution immune to detector attacks},\ }\href
  {https://doi.org/10.1103/PhysRevLett.111.130501} {\bibfield  {journal}
  {\bibinfo  {journal} {Phys. Rev. Lett.}\ }\textbf {\bibinfo {volume} {111}},\
  \bibinfo {pages} {130501} (\bibinfo {year} {2013})}\BibitemShut {NoStop}%
\bibitem [{\citenamefont {Bersin}\ \emph {et~al.}(2024)\citenamefont {Bersin},
  \citenamefont {Grein}, \citenamefont {Sutula}, \citenamefont {Murphy},
  \citenamefont {Huan}, \citenamefont {Stevens}, \citenamefont {Suleymanzade},
  \citenamefont {Lee}, \citenamefont {Riedinger}, \citenamefont {Starling},
  \citenamefont {Stas}, \citenamefont {Knaut}, \citenamefont {Sinclair},
  \citenamefont {Assumpcao}, \citenamefont {Wei}, \citenamefont {Knall},
  \citenamefont {Machielse}, \citenamefont {Sukachev}, \citenamefont
  {Levonian}, \citenamefont {Bhaskar}, \citenamefont
  {Lon\ifmmode~\check{c}\else \v{c}\fi{}ar}, \citenamefont {Hamilton},
  \citenamefont {Lukin}, \citenamefont {Englund},\ and\ \citenamefont
  {Dixon}}]{BostonQuantumNetwork}%
  \BibitemOpen
  \bibfield  {author} {\bibinfo {author} {\bibfnamefont {E.}~\bibnamefont
  {Bersin}}, \bibinfo {author} {\bibfnamefont {M.}~\bibnamefont {Grein}},
  \bibinfo {author} {\bibfnamefont {M.}~\bibnamefont {Sutula}}, \bibinfo
  {author} {\bibfnamefont {R.}~\bibnamefont {Murphy}}, \bibinfo {author}
  {\bibfnamefont {Y.~Q.}\ \bibnamefont {Huan}}, \bibinfo {author}
  {\bibfnamefont {M.}~\bibnamefont {Stevens}}, \bibinfo {author} {\bibfnamefont
  {A.}~\bibnamefont {Suleymanzade}}, \bibinfo {author} {\bibfnamefont
  {C.}~\bibnamefont {Lee}}, \bibinfo {author} {\bibfnamefont {R.}~\bibnamefont
  {Riedinger}}, \bibinfo {author} {\bibfnamefont {D.~J.}\ \bibnamefont
  {Starling}}, \bibinfo {author} {\bibfnamefont {P.-J.}\ \bibnamefont {Stas}},
  \bibinfo {author} {\bibfnamefont {C.~M.}\ \bibnamefont {Knaut}}, \bibinfo
  {author} {\bibfnamefont {N.}~\bibnamefont {Sinclair}}, \bibinfo {author}
  {\bibfnamefont {D.~R.}\ \bibnamefont {Assumpcao}}, \bibinfo {author}
  {\bibfnamefont {Y.-C.}\ \bibnamefont {Wei}}, \bibinfo {author} {\bibfnamefont
  {E.~N.}\ \bibnamefont {Knall}}, \bibinfo {author} {\bibfnamefont
  {B.}~\bibnamefont {Machielse}}, \bibinfo {author} {\bibfnamefont {D.~D.}\
  \bibnamefont {Sukachev}}, \bibinfo {author} {\bibfnamefont {D.~S.}\
  \bibnamefont {Levonian}}, \bibinfo {author} {\bibfnamefont {M.~K.}\
  \bibnamefont {Bhaskar}}, \bibinfo {author} {\bibfnamefont {M.}~\bibnamefont
  {Lon\ifmmode~\check{c}\else \v{c}\fi{}ar}}, \bibinfo {author} {\bibfnamefont
  {S.}~\bibnamefont {Hamilton}}, \bibinfo {author} {\bibfnamefont
  {M.}~\bibnamefont {Lukin}}, \bibinfo {author} {\bibfnamefont
  {D.}~\bibnamefont {Englund}},\ and\ \bibinfo {author} {\bibfnamefont {P.~B.}\
  \bibnamefont {Dixon}},\ }\bibfield  {title} {\bibinfo {title} {Development of
  a boston-area 50-km fiber quantum network testbed},\ }\href
  {https://doi.org/10.1103/PhysRevApplied.21.014024} {\bibfield  {journal}
  {\bibinfo  {journal} {Phys. Rev. Appl.}\ }\textbf {\bibinfo {volume} {21}},\
  \bibinfo {pages} {014024} (\bibinfo {year} {2024})}\BibitemShut {NoStop}%
\bibitem [{\citenamefont {Pirandola}\ \emph {et~al.}(2015)\citenamefont
  {Pirandola}, \citenamefont {Ottaviani}, \citenamefont {Spedalieri},
  \citenamefont {Weedbrook}, \citenamefont {Braunstein}, \citenamefont {Lloyd},
  \citenamefont {Gehring}, \citenamefont {Jacobsen},\ and\ \citenamefont
  {Andersen}}]{pirandola2015}%
  \BibitemOpen
  \bibfield  {author} {\bibinfo {author} {\bibfnamefont {S.}~\bibnamefont
  {Pirandola}}, \bibinfo {author} {\bibfnamefont {C.}~\bibnamefont
  {Ottaviani}}, \bibinfo {author} {\bibfnamefont {G.}~\bibnamefont
  {Spedalieri}}, \bibinfo {author} {\bibfnamefont {C.}~\bibnamefont
  {Weedbrook}}, \bibinfo {author} {\bibfnamefont {S.~L.}\ \bibnamefont
  {Braunstein}}, \bibinfo {author} {\bibfnamefont {S.}~\bibnamefont {Lloyd}},
  \bibinfo {author} {\bibfnamefont {T.}~\bibnamefont {Gehring}}, \bibinfo
  {author} {\bibfnamefont {C.~S.}\ \bibnamefont {Jacobsen}},\ and\ \bibinfo
  {author} {\bibfnamefont {U.~L.}\ \bibnamefont {Andersen}},\ }\bibfield
  {title} {\bibinfo {title} {High-rate measurement-device-independent quantum
  cryptography},\ }\href {https://doi.org/10.1038/nphoton.2015.83} {\bibfield
  {journal} {\bibinfo  {journal} {Nat. Photonics}\ }\textbf {\bibinfo {volume}
  {9}},\ \bibinfo {pages} {397} (\bibinfo {year} {2015})}\BibitemShut {NoStop}%
\bibitem [{\citenamefont {Valivarthi}\ \emph {et~al.}(2016)\citenamefont
  {Valivarthi}, \citenamefont {Grimau~Puigibert}, \citenamefont {Zhou},
  \citenamefont {Aguilar}, \citenamefont {Verma}, \citenamefont {Marsili},
  \citenamefont {Shaw}, \citenamefont {Nam}, \citenamefont {Oblak},\ and\
  \citenamefont {Tittel}}]{Valivarthi2016}%
  \BibitemOpen
  \bibfield  {author} {\bibinfo {author} {\bibfnamefont {R.}~\bibnamefont
  {Valivarthi}}, \bibinfo {author} {\bibfnamefont {M.}~\bibnamefont
  {Grimau~Puigibert}}, \bibinfo {author} {\bibfnamefont {Q.}~\bibnamefont
  {Zhou}}, \bibinfo {author} {\bibfnamefont {G.~H.}\ \bibnamefont {Aguilar}},
  \bibinfo {author} {\bibfnamefont {V.~B.}\ \bibnamefont {Verma}}, \bibinfo
  {author} {\bibfnamefont {F.}~\bibnamefont {Marsili}}, \bibinfo {author}
  {\bibfnamefont {M.~D.}\ \bibnamefont {Shaw}}, \bibinfo {author}
  {\bibfnamefont {S.~W.}\ \bibnamefont {Nam}}, \bibinfo {author} {\bibfnamefont
  {D.}~\bibnamefont {Oblak}},\ and\ \bibinfo {author} {\bibfnamefont
  {W.}~\bibnamefont {Tittel}},\ }\bibfield  {title} {\bibinfo {title} {Quantum
  teleportation across a metropolitan fibre network},\ }\href
  {https://doi.org/10.1038/nphoton.2016.180} {\bibfield  {journal} {\bibinfo
  {journal} {Nature Photonics}\ }\textbf {\bibinfo {volume} {10}},\ \bibinfo
  {pages} {676} (\bibinfo {year} {2016})}\BibitemShut {NoStop}%
\bibitem [{\citenamefont {Craddock}\ \emph {et~al.}(2024)\citenamefont
  {Craddock}, \citenamefont {Lazenby}, \citenamefont {Portmann}, \citenamefont
  {Sekelsky}, \citenamefont {Flament},\ and\ \citenamefont
  {Namazi}}]{Craddock2024}%
  \BibitemOpen
  \bibfield  {author} {\bibinfo {author} {\bibfnamefont {A.~N.}\ \bibnamefont
  {Craddock}}, \bibinfo {author} {\bibfnamefont {A.}~\bibnamefont {Lazenby}},
  \bibinfo {author} {\bibfnamefont {G.~B.}\ \bibnamefont {Portmann}}, \bibinfo
  {author} {\bibfnamefont {R.}~\bibnamefont {Sekelsky}}, \bibinfo {author}
  {\bibfnamefont {M.}~\bibnamefont {Flament}},\ and\ \bibinfo {author}
  {\bibfnamefont {M.}~\bibnamefont {Namazi}},\ }\bibfield  {title} {\bibinfo
  {title} {Automated distribution of polarization-entangled photons using
  deployed new york city fibers},\ }\href
  {https://doi.org/10.1103/PRXQuantum.5.030330} {\bibfield  {journal} {\bibinfo
   {journal} {PRX Quantum}\ }\textbf {\bibinfo {volume} {5}},\ \bibinfo {pages}
  {030330} (\bibinfo {year} {2024})}\BibitemShut {NoStop}%
\bibitem [{\citenamefont {Tang}\ \emph {et~al.}(2016)\citenamefont {Tang},
  \citenamefont {Yin}, \citenamefont {Zhao}, \citenamefont {Liu}, \citenamefont
  {Sun}, \citenamefont {Huang}, \citenamefont {Zhang}, \citenamefont {Chen},
  \citenamefont {Zhang}, \citenamefont {You}, \citenamefont {Wang},
  \citenamefont {Liu}, \citenamefont {Lu}, \citenamefont {Jiang}, \citenamefont
  {Ma}, \citenamefont {Zhang}, \citenamefont {Chen},\ and\ \citenamefont
  {Pan}}]{JianWeiPan2016MDIQKD}%
  \BibitemOpen
  \bibfield  {author} {\bibinfo {author} {\bibfnamefont {Y.-L.}\ \bibnamefont
  {Tang}}, \bibinfo {author} {\bibfnamefont {H.-L.}\ \bibnamefont {Yin}},
  \bibinfo {author} {\bibfnamefont {Q.}~\bibnamefont {Zhao}}, \bibinfo {author}
  {\bibfnamefont {H.}~\bibnamefont {Liu}}, \bibinfo {author} {\bibfnamefont
  {X.-X.}\ \bibnamefont {Sun}}, \bibinfo {author} {\bibfnamefont {M.-Q.}\
  \bibnamefont {Huang}}, \bibinfo {author} {\bibfnamefont {W.-J.}\ \bibnamefont
  {Zhang}}, \bibinfo {author} {\bibfnamefont {S.-J.}\ \bibnamefont {Chen}},
  \bibinfo {author} {\bibfnamefont {L.}~\bibnamefont {Zhang}}, \bibinfo
  {author} {\bibfnamefont {L.-X.}\ \bibnamefont {You}}, \bibinfo {author}
  {\bibfnamefont {Z.}~\bibnamefont {Wang}}, \bibinfo {author} {\bibfnamefont
  {Y.}~\bibnamefont {Liu}}, \bibinfo {author} {\bibfnamefont {C.-Y.}\
  \bibnamefont {Lu}}, \bibinfo {author} {\bibfnamefont {X.}~\bibnamefont
  {Jiang}}, \bibinfo {author} {\bibfnamefont {X.}~\bibnamefont {Ma}}, \bibinfo
  {author} {\bibfnamefont {Q.}~\bibnamefont {Zhang}}, \bibinfo {author}
  {\bibfnamefont {T.-Y.}\ \bibnamefont {Chen}},\ and\ \bibinfo {author}
  {\bibfnamefont {J.-W.}\ \bibnamefont {Pan}},\ }\bibfield  {title} {\bibinfo
  {title} {Measurement-device-independent quantum key distribution over
  untrustful metropolitan network},\ }\href
  {https://doi.org/10.1103/PhysRevX.6.011024} {\bibfield  {journal} {\bibinfo
  {journal} {Phys. Rev. X}\ }\textbf {\bibinfo {volume} {6}},\ \bibinfo {pages}
  {011024} (\bibinfo {year} {2016})}\BibitemShut {NoStop}%
\bibitem [{\citenamefont {Cao}\ \emph {et~al.}(2020)\citenamefont {Cao},
  \citenamefont {Li}, \citenamefont {Yang}, \citenamefont {Jiang},
  \citenamefont {Li}, \citenamefont {Hu}, \citenamefont {Abulizi},
  \citenamefont {Li}, \citenamefont {Zhang}, \citenamefont {Sun}, \citenamefont
  {Liu}, \citenamefont {Jiang}, \citenamefont {Liao}, \citenamefont {Ren},
  \citenamefont {Li}, \citenamefont {You}, \citenamefont {Wang}, \citenamefont
  {Yin}, \citenamefont {Lu}, \citenamefont {Wang}, \citenamefont {Zhang},
  \citenamefont {Peng},\ and\ \citenamefont {Pan}}]{JianWeiPan2020}%
  \BibitemOpen
  \bibfield  {author} {\bibinfo {author} {\bibfnamefont {Y.}~\bibnamefont
  {Cao}}, \bibinfo {author} {\bibfnamefont {Y.-H.}\ \bibnamefont {Li}},
  \bibinfo {author} {\bibfnamefont {K.-X.}\ \bibnamefont {Yang}}, \bibinfo
  {author} {\bibfnamefont {Y.-F.}\ \bibnamefont {Jiang}}, \bibinfo {author}
  {\bibfnamefont {S.-L.}\ \bibnamefont {Li}}, \bibinfo {author} {\bibfnamefont
  {X.-L.}\ \bibnamefont {Hu}}, \bibinfo {author} {\bibfnamefont
  {M.}~\bibnamefont {Abulizi}}, \bibinfo {author} {\bibfnamefont {C.-L.}\
  \bibnamefont {Li}}, \bibinfo {author} {\bibfnamefont {W.}~\bibnamefont
  {Zhang}}, \bibinfo {author} {\bibfnamefont {Q.-C.}\ \bibnamefont {Sun}},
  \bibinfo {author} {\bibfnamefont {W.-Y.}\ \bibnamefont {Liu}}, \bibinfo
  {author} {\bibfnamefont {X.}~\bibnamefont {Jiang}}, \bibinfo {author}
  {\bibfnamefont {S.-K.}\ \bibnamefont {Liao}}, \bibinfo {author}
  {\bibfnamefont {J.-G.}\ \bibnamefont {Ren}}, \bibinfo {author} {\bibfnamefont
  {H.}~\bibnamefont {Li}}, \bibinfo {author} {\bibfnamefont {L.}~\bibnamefont
  {You}}, \bibinfo {author} {\bibfnamefont {Z.}~\bibnamefont {Wang}}, \bibinfo
  {author} {\bibfnamefont {J.}~\bibnamefont {Yin}}, \bibinfo {author}
  {\bibfnamefont {C.-Y.}\ \bibnamefont {Lu}}, \bibinfo {author} {\bibfnamefont
  {X.-B.}\ \bibnamefont {Wang}}, \bibinfo {author} {\bibfnamefont
  {Q.}~\bibnamefont {Zhang}}, \bibinfo {author} {\bibfnamefont {C.-Z.}\
  \bibnamefont {Peng}},\ and\ \bibinfo {author} {\bibfnamefont {J.-W.}\
  \bibnamefont {Pan}},\ }\bibfield  {title} {\bibinfo {title} {Long-distance
  free-space measurement-device-independent quantum key distribution},\ }\href
  {https://doi.org/10.1103/PhysRevLett.125.260503} {\bibfield  {journal}
  {\bibinfo  {journal} {Phys. Rev. Lett.}\ }\textbf {\bibinfo {volume} {125}},\
  \bibinfo {pages} {260503} (\bibinfo {year} {2020})}\BibitemShut {NoStop}%
\bibitem [{\citenamefont {Bhaskar}\ \emph {et~al.}(2020)\citenamefont
  {Bhaskar}, \citenamefont {Riedinger}, \citenamefont {Machielse},
  \citenamefont {Levonian}, \citenamefont {Nguyen}, \citenamefont {Knall},
  \citenamefont {Park}, \citenamefont {Englund}, \citenamefont {Lončar},
  \citenamefont {Sukachev},\ and\ \citenamefont {Lukin}}]{Bhaskar2020}%
  \BibitemOpen
  \bibfield  {author} {\bibinfo {author} {\bibfnamefont {M.~K.}\ \bibnamefont
  {Bhaskar}}, \bibinfo {author} {\bibfnamefont {R.}~\bibnamefont {Riedinger}},
  \bibinfo {author} {\bibfnamefont {B.}~\bibnamefont {Machielse}}, \bibinfo
  {author} {\bibfnamefont {D.~S.}\ \bibnamefont {Levonian}}, \bibinfo {author}
  {\bibfnamefont {C.~T.}\ \bibnamefont {Nguyen}}, \bibinfo {author}
  {\bibfnamefont {E.~N.}\ \bibnamefont {Knall}}, \bibinfo {author}
  {\bibfnamefont {H.}~\bibnamefont {Park}}, \bibinfo {author} {\bibfnamefont
  {D.}~\bibnamefont {Englund}}, \bibinfo {author} {\bibfnamefont
  {M.}~\bibnamefont {Lončar}}, \bibinfo {author} {\bibfnamefont {D.~D.}\
  \bibnamefont {Sukachev}},\ and\ \bibinfo {author} {\bibfnamefont {M.~D.}\
  \bibnamefont {Lukin}},\ }\bibfield  {title} {\bibinfo {title} {Experimental
  demonstration of memory-enhanced quantum communication},\ }\href
  {https://doi.org/10.1038/s41586-020-2103-5} {\bibfield  {journal} {\bibinfo
  {journal} {Nature}\ }\textbf {\bibinfo {volume} {580}},\ \bibinfo {pages}
  {60–64} (\bibinfo {year} {2020})}\BibitemShut {NoStop}%
\bibitem [{\citenamefont {Bunandar}\ \emph {et~al.}(2018)\citenamefont
  {Bunandar}, \citenamefont {Lentine}, \citenamefont {Lee}, \citenamefont
  {Cai}, \citenamefont {Long}, \citenamefont {Boynton}, \citenamefont
  {Martinez}, \citenamefont {DeRose}, \citenamefont {Chen}, \citenamefont
  {Grein}, \citenamefont {Trotter}, \citenamefont {Starbuck}, \citenamefont
  {Pomerene}, \citenamefont {Hamilton}, \citenamefont {Wong}, \citenamefont
  {Camacho}, \citenamefont {Davids}, \citenamefont {Urayama},\ and\
  \citenamefont {Englund}}]{BostonQKD2018}%
  \BibitemOpen
  \bibfield  {author} {\bibinfo {author} {\bibfnamefont {D.}~\bibnamefont
  {Bunandar}}, \bibinfo {author} {\bibfnamefont {A.}~\bibnamefont {Lentine}},
  \bibinfo {author} {\bibfnamefont {C.}~\bibnamefont {Lee}}, \bibinfo {author}
  {\bibfnamefont {H.}~\bibnamefont {Cai}}, \bibinfo {author} {\bibfnamefont
  {C.~M.}\ \bibnamefont {Long}}, \bibinfo {author} {\bibfnamefont
  {N.}~\bibnamefont {Boynton}}, \bibinfo {author} {\bibfnamefont
  {N.}~\bibnamefont {Martinez}}, \bibinfo {author} {\bibfnamefont
  {C.}~\bibnamefont {DeRose}}, \bibinfo {author} {\bibfnamefont
  {C.}~\bibnamefont {Chen}}, \bibinfo {author} {\bibfnamefont {M.}~\bibnamefont
  {Grein}}, \bibinfo {author} {\bibfnamefont {D.}~\bibnamefont {Trotter}},
  \bibinfo {author} {\bibfnamefont {A.}~\bibnamefont {Starbuck}}, \bibinfo
  {author} {\bibfnamefont {A.}~\bibnamefont {Pomerene}}, \bibinfo {author}
  {\bibfnamefont {S.}~\bibnamefont {Hamilton}}, \bibinfo {author}
  {\bibfnamefont {F.~N.}\ \bibnamefont {Wong}}, \bibinfo {author}
  {\bibfnamefont {R.}~\bibnamefont {Camacho}}, \bibinfo {author} {\bibfnamefont
  {P.}~\bibnamefont {Davids}}, \bibinfo {author} {\bibfnamefont
  {J.}~\bibnamefont {Urayama}},\ and\ \bibinfo {author} {\bibfnamefont
  {D.}~\bibnamefont {Englund}},\ }\bibfield  {title} {\bibinfo {title}
  {Metropolitan quantum key distribution with silicon photonics},\ }\bibfield
  {journal} {\bibinfo  {journal} {Physical Review X}\ }\textbf {\bibinfo
  {volume} {8}},\ \href {https://doi.org/10.1103/physrevx.8.021009}
  {10.1103/physrevx.8.021009} (\bibinfo {year} {2018})\BibitemShut {NoStop}%
\bibitem [{\citenamefont {Sit}\ \emph {et~al.}(2017)\citenamefont {Sit},
  \citenamefont {Bouchard}, \citenamefont {Fickler}, \citenamefont
  {Gagnon-Bischoff}, \citenamefont {Larocque}, \citenamefont {Heshami},
  \citenamefont {Elser}, \citenamefont {Peuntinger}, \citenamefont {Günthner},
  \citenamefont {Heim}, \citenamefont {Marquardt}, \citenamefont {Leuchs},
  \citenamefont {Boyd},\ and\ \citenamefont {Karimi}}]{Sit2017}%
  \BibitemOpen
  \bibfield  {author} {\bibinfo {author} {\bibfnamefont {A.}~\bibnamefont
  {Sit}}, \bibinfo {author} {\bibfnamefont {F.}~\bibnamefont {Bouchard}},
  \bibinfo {author} {\bibfnamefont {R.}~\bibnamefont {Fickler}}, \bibinfo
  {author} {\bibfnamefont {J.}~\bibnamefont {Gagnon-Bischoff}}, \bibinfo
  {author} {\bibfnamefont {H.}~\bibnamefont {Larocque}}, \bibinfo {author}
  {\bibfnamefont {K.}~\bibnamefont {Heshami}}, \bibinfo {author} {\bibfnamefont
  {D.}~\bibnamefont {Elser}}, \bibinfo {author} {\bibfnamefont
  {C.}~\bibnamefont {Peuntinger}}, \bibinfo {author} {\bibfnamefont
  {K.}~\bibnamefont {Günthner}}, \bibinfo {author} {\bibfnamefont
  {B.}~\bibnamefont {Heim}}, \bibinfo {author} {\bibfnamefont {C.}~\bibnamefont
  {Marquardt}}, \bibinfo {author} {\bibfnamefont {G.}~\bibnamefont {Leuchs}},
  \bibinfo {author} {\bibfnamefont {R.~W.}\ \bibnamefont {Boyd}},\ and\
  \bibinfo {author} {\bibfnamefont {E.}~\bibnamefont {Karimi}},\ }\bibfield
  {title} {\bibinfo {title} {High-dimensional intracity quantum cryptography
  with structured photons},\ }\href {https://doi.org/10.1364/optica.4.001006}
  {\bibfield  {journal} {\bibinfo  {journal} {Optica}\ }\textbf {\bibinfo
  {volume} {4}},\ \bibinfo {pages} {1006} (\bibinfo {year} {2017})}\BibitemShut
  {NoStop}%
\bibitem [{\citenamefont {Pugh}\ \emph {et~al.}(2017)\citenamefont {Pugh},
  \citenamefont {Kaiser}, \citenamefont {Bourgoin}, \citenamefont {Jin},
  \citenamefont {Sultana}, \citenamefont {Agne}, \citenamefont {Anisimova},
  \citenamefont {Makarov}, \citenamefont {Choi}, \citenamefont {Higgins},\ and\
  \citenamefont {Jennewein}}]{Pugh2017}%
  \BibitemOpen
  \bibfield  {author} {\bibinfo {author} {\bibfnamefont {C.~J.}\ \bibnamefont
  {Pugh}}, \bibinfo {author} {\bibfnamefont {S.}~\bibnamefont {Kaiser}},
  \bibinfo {author} {\bibfnamefont {J.-P.}\ \bibnamefont {Bourgoin}}, \bibinfo
  {author} {\bibfnamefont {J.}~\bibnamefont {Jin}}, \bibinfo {author}
  {\bibfnamefont {N.}~\bibnamefont {Sultana}}, \bibinfo {author} {\bibfnamefont
  {S.}~\bibnamefont {Agne}}, \bibinfo {author} {\bibfnamefont {E.}~\bibnamefont
  {Anisimova}}, \bibinfo {author} {\bibfnamefont {V.}~\bibnamefont {Makarov}},
  \bibinfo {author} {\bibfnamefont {E.}~\bibnamefont {Choi}}, \bibinfo {author}
  {\bibfnamefont {B.~L.}\ \bibnamefont {Higgins}},\ and\ \bibinfo {author}
  {\bibfnamefont {T.}~\bibnamefont {Jennewein}},\ }\bibfield  {title} {\bibinfo
  {title} {Airborne demonstration of a quantum key distribution receiver
  payload},\ }\href {https://doi.org/10.1088/2058-9565/aa701f} {\bibfield
  {journal} {\bibinfo  {journal} {Quantum Science and Technology}\ }\textbf
  {\bibinfo {volume} {2}},\ \bibinfo {pages} {024009} (\bibinfo {year}
  {2017})}\BibitemShut {NoStop}%
\bibitem [{\citenamefont {Brougham}\ and\ \citenamefont
  {Oi}(2022)}]{Brougham2022}%
  \BibitemOpen
  \bibfield  {author} {\bibinfo {author} {\bibfnamefont {T.}~\bibnamefont
  {Brougham}}\ and\ \bibinfo {author} {\bibfnamefont {D.~K.~L.}\ \bibnamefont
  {Oi}},\ }\bibfield  {title} {\bibinfo {title} {Modelling efficient bb84 with
  applications for medium-range, terrestrial free-space qkd},\ }\href
  {https://doi.org/10.1088/1367-2630/ac7f4e} {\bibfield  {journal} {\bibinfo
  {journal} {New Journal of Physics}\ }\textbf {\bibinfo {volume} {24}},\
  \bibinfo {pages} {075002} (\bibinfo {year} {2022})}\BibitemShut {NoStop}%
\bibitem [{\citenamefont {Ghalaii}\ and\ \citenamefont
  {Pirandola}(2023)}]{Ghalaii2023}%
  \BibitemOpen
  \bibfield  {author} {\bibinfo {author} {\bibfnamefont {M.}~\bibnamefont
  {Ghalaii}}\ and\ \bibinfo {author} {\bibfnamefont {S.}~\bibnamefont
  {Pirandola}},\ }\bibfield  {title} {\bibinfo {title} {Continuous-variable
  measurement-device-independent quantum key distribution in free-space
  channels},\ }\href {https://doi.org/10.1103/PhysRevA.108.042621} {\bibfield
  {journal} {\bibinfo  {journal} {Phys. Rev. A}\ }\textbf {\bibinfo {volume}
  {108}},\ \bibinfo {pages} {042621} (\bibinfo {year} {2023})}\BibitemShut
  {NoStop}%
\bibitem [{\citenamefont {Jaouni}\ \emph {et~al.}(2025)\citenamefont {Jaouni},
  \citenamefont {Scarfe}, \citenamefont {Bouchard}, \citenamefont {Krenn},
  \citenamefont {Heshami}, \citenamefont {Di~Colandrea},\ and\ \citenamefont
  {Karimi}}]{Jaouni2025}%
  \BibitemOpen
  \bibfield  {author} {\bibinfo {author} {\bibfnamefont {T.}~\bibnamefont
  {Jaouni}}, \bibinfo {author} {\bibfnamefont {L.}~\bibnamefont {Scarfe}},
  \bibinfo {author} {\bibfnamefont {F.}~\bibnamefont {Bouchard}}, \bibinfo
  {author} {\bibfnamefont {M.}~\bibnamefont {Krenn}}, \bibinfo {author}
  {\bibfnamefont {K.}~\bibnamefont {Heshami}}, \bibinfo {author} {\bibfnamefont
  {F.}~\bibnamefont {Di~Colandrea}},\ and\ \bibinfo {author} {\bibfnamefont
  {E.}~\bibnamefont {Karimi}},\ }\bibfield  {title} {\bibinfo {title}
  {Predicting atmospheric turbulence for secure quantum communications in free
  space},\ }\href {https://doi.org/10.1364/oe.546606} {\bibfield  {journal}
  {\bibinfo  {journal} {Optics Express}\ }\textbf {\bibinfo {volume} {33}},\
  \bibinfo {pages} {10759} (\bibinfo {year} {2025})}\BibitemShut {NoStop}%
\bibitem [{\citenamefont {Scarfe}\ \emph {et~al.}(2025)\citenamefont {Scarfe},
  \citenamefont {Hufnagel}, \citenamefont {Ferrer-Garcia}, \citenamefont
  {D’Errico}, \citenamefont {Heshami},\ and\ \citenamefont
  {Karimi}}]{Scarfe2025}%
  \BibitemOpen
  \bibfield  {author} {\bibinfo {author} {\bibfnamefont {L.}~\bibnamefont
  {Scarfe}}, \bibinfo {author} {\bibfnamefont {F.}~\bibnamefont {Hufnagel}},
  \bibinfo {author} {\bibfnamefont {M.~F.}\ \bibnamefont {Ferrer-Garcia}},
  \bibinfo {author} {\bibfnamefont {A.}~\bibnamefont {D’Errico}}, \bibinfo
  {author} {\bibfnamefont {K.}~\bibnamefont {Heshami}},\ and\ \bibinfo {author}
  {\bibfnamefont {E.}~\bibnamefont {Karimi}},\ }\bibfield  {title} {\bibinfo
  {title} {Fast adaptive optics for high-dimensional quantum communications in
  turbulent channels},\ }\href {https://doi.org/10.1038/s42005-025-01986-6}
  {\bibfield  {journal} {\bibinfo  {journal} {Communications Physics}\ }\textbf
  {\bibinfo {volume} {8}},\ \bibinfo {pages} {79} (\bibinfo {year}
  {2025})}\BibitemShut {NoStop}%
\bibitem [{\citenamefont {Wang}\ \emph
  {et~al.}(2022{\natexlab{a}})\citenamefont {Wang}, \citenamefont {Yin},
  \citenamefont {He}, \citenamefont {Chen}, \citenamefont {Wang}, \citenamefont
  {Ye}, \citenamefont {Zhou}, \citenamefont {Fan-Yuan}, \citenamefont {Wang},
  \citenamefont {Chen}, \citenamefont {Zhu}, \citenamefont {Morozov},
  \citenamefont {Divochiy}, \citenamefont {Zhou}, \citenamefont {Guo},\ and\
  \citenamefont {Han}}]{WangTF2022}%
  \BibitemOpen
  \bibfield  {author} {\bibinfo {author} {\bibfnamefont {S.}~\bibnamefont
  {Wang}}, \bibinfo {author} {\bibfnamefont {Z.-Q.}\ \bibnamefont {Yin}},
  \bibinfo {author} {\bibfnamefont {D.-Y.}\ \bibnamefont {He}}, \bibinfo
  {author} {\bibfnamefont {W.}~\bibnamefont {Chen}}, \bibinfo {author}
  {\bibfnamefont {R.-Q.}\ \bibnamefont {Wang}}, \bibinfo {author}
  {\bibfnamefont {P.}~\bibnamefont {Ye}}, \bibinfo {author} {\bibfnamefont
  {Y.}~\bibnamefont {Zhou}}, \bibinfo {author} {\bibfnamefont {G.-J.}\
  \bibnamefont {Fan-Yuan}}, \bibinfo {author} {\bibfnamefont {F.-X.}\
  \bibnamefont {Wang}}, \bibinfo {author} {\bibfnamefont {W.}~\bibnamefont
  {Chen}}, \bibinfo {author} {\bibfnamefont {Y.-G.}\ \bibnamefont {Zhu}},
  \bibinfo {author} {\bibfnamefont {P.~V.}\ \bibnamefont {Morozov}}, \bibinfo
  {author} {\bibfnamefont {A.~V.}\ \bibnamefont {Divochiy}}, \bibinfo {author}
  {\bibfnamefont {Z.}~\bibnamefont {Zhou}}, \bibinfo {author} {\bibfnamefont
  {G.-C.}\ \bibnamefont {Guo}},\ and\ \bibinfo {author} {\bibfnamefont {Z.-F.}\
  \bibnamefont {Han}},\ }\bibfield  {title} {\bibinfo {title} {Twin-field
  quantum key distribution over 830-km fibre},\ }\href
  {https://doi.org/10.1038/s41566-021-00928-2} {\bibfield  {journal} {\bibinfo
  {journal} {Nature Photonics}\ }\textbf {\bibinfo {volume} {16}},\ \bibinfo
  {pages} {154} (\bibinfo {year} {2022}{\natexlab{a}})}\BibitemShut {NoStop}%
\bibitem [{\citenamefont {Wang}\ and\ \citenamefont
  {Wang}(2014)}]{Wang2014Simulating}%
  \BibitemOpen
  \bibfield  {author} {\bibinfo {author} {\bibfnamefont {Q.}~\bibnamefont
  {Wang}}\ and\ \bibinfo {author} {\bibfnamefont {X.-B.}\ \bibnamefont
  {Wang}},\ }\bibfield  {title} {\bibinfo {title} {Simulating of the
  measurement-device independent quantum key distribution with phase randomized
  general sources},\ }\href {https://doi.org/10.1038/srep04612} {\bibfield
  {journal} {\bibinfo  {journal} {Scientific Reports}\ }\textbf {\bibinfo
  {volume} {4}},\ \bibinfo {pages} {4612} (\bibinfo {year} {2014})}\BibitemShut
  {NoStop}%
\bibitem [{\citenamefont {Bel}\ and\ \citenamefont {Kiran}(2025)}]{Bel2025}%
  \BibitemOpen
  \bibfield  {author} {\bibinfo {author} {\bibfnamefont {O.}~\bibnamefont
  {Bel}}\ and\ \bibinfo {author} {\bibfnamefont {M.}~\bibnamefont {Kiran}},\
  }\bibfield  {title} {\bibinfo {title} {Simulators for quantum network
  modeling: A comprehensive review},\ }\href
  {https://doi.org/10.1016/j.comnet.2025.111204} {\bibfield  {journal}
  {\bibinfo  {journal} {Computer Networks}\ }\textbf {\bibinfo {volume}
  {263}},\ \bibinfo {pages} {111204} (\bibinfo {year} {2025})}\BibitemShut
  {NoStop}%
\bibitem [{\citenamefont {Panayi}\ \emph {et~al.}(2014)\citenamefont {Panayi},
  \citenamefont {Razavi}, \citenamefont {Ma},\ and\ \citenamefont
  {L{\"u}tkenhaus}}]{Panayi2014}%
  \BibitemOpen
  \bibfield  {author} {\bibinfo {author} {\bibfnamefont {C.}~\bibnamefont
  {Panayi}}, \bibinfo {author} {\bibfnamefont {M.}~\bibnamefont {Razavi}},
  \bibinfo {author} {\bibfnamefont {X.}~\bibnamefont {Ma}},\ and\ \bibinfo
  {author} {\bibfnamefont {N.}~\bibnamefont {L{\"u}tkenhaus}},\ }\bibfield
  {title} {\bibinfo {title} {Memory-assisted measurement-device-independent
  quantum key distribution},\ }\href
  {https://doi.org/10.1088/1367-2630/16/4/043005} {\bibfield  {journal}
  {\bibinfo  {journal} {New Journal of Physics}\ }\textbf {\bibinfo {volume}
  {16}},\ \bibinfo {pages} {043005} (\bibinfo {year} {2014})},\ \Eprint
  {https://arxiv.org/abs/1309.3406} {arXiv:1309.3406} \BibitemShut {NoStop}%
\bibitem [{\citenamefont {Sangouard}\ \emph {et~al.}(2011)\citenamefont
  {Sangouard}, \citenamefont {Simon}, \citenamefont {de~Riedmatten},\ and\
  \citenamefont {Gisin}}]{Simon2011}%
  \BibitemOpen
  \bibfield  {author} {\bibinfo {author} {\bibfnamefont {N.}~\bibnamefont
  {Sangouard}}, \bibinfo {author} {\bibfnamefont {C.}~\bibnamefont {Simon}},
  \bibinfo {author} {\bibfnamefont {H.}~\bibnamefont {de~Riedmatten}},\ and\
  \bibinfo {author} {\bibfnamefont {N.}~\bibnamefont {Gisin}},\ }\bibfield
  {title} {\bibinfo {title} {Quantum repeaters based on atomic ensembles and
  linear optics},\ }\href {https://doi.org/10.1103/RevModPhys.83.33} {\bibfield
   {journal} {\bibinfo  {journal} {Rev. Mod. Phys.}\ }\textbf {\bibinfo
  {volume} {83}},\ \bibinfo {pages} {33} (\bibinfo {year} {2011})}\BibitemShut
  {NoStop}%
\bibitem [{\citenamefont {Rajpurohit}\ \emph {et~al.}(2024)\citenamefont
  {Rajpurohit}, \citenamefont {Deka}, \citenamefont {Singh},\ and\
  \citenamefont {P}}]{AdvancedQKD2024}%
  \BibitemOpen
  \bibfield  {author} {\bibinfo {author} {\bibfnamefont {M.}~\bibnamefont
  {Rajpurohit}}, \bibinfo {author} {\bibfnamefont {K.~M.}\ \bibnamefont
  {Deka}}, \bibinfo {author} {\bibfnamefont {V.~P.}\ \bibnamefont {Singh}},\
  and\ \bibinfo {author} {\bibfnamefont {H.}~\bibnamefont {P}},\ }\bibfield
  {title} {\bibinfo {title} {Advanced qkd protocols and practical challenges},\
  }in\ \href {https://doi.org/10.1109/PKIA62599.2024.10727858} {\emph {\bibinfo
  {booktitle} {2024 IEEE International Conference on Public Key Infrastructure
  and its Applications (PKIA)}}}\ (\bibinfo {year} {2024})\ pp.\ \bibinfo
  {pages} {1--8}\BibitemShut {NoStop}%
\bibitem [{\citenamefont {Kržič}\ \emph {et~al.}(2023)\citenamefont
  {Kržič}, \citenamefont {Sharma}, \citenamefont {Spiess}, \citenamefont
  {Chandrashekara}, \citenamefont {Töpfer}, \citenamefont {Sauer},
  \citenamefont {González-Martín~del Campo}, \citenamefont {Kopf},
  \citenamefont {Petscharnig}, \citenamefont {Grafenauer}, \citenamefont
  {Lieger}, \citenamefont {Ömer}, \citenamefont {Pacher}, \citenamefont
  {Berlich}, \citenamefont {Peschel}, \citenamefont {Damm}, \citenamefont
  {Risse}, \citenamefont {Goy}, \citenamefont {Rieländer}, \citenamefont
  {Tünnermann},\ and\ \citenamefont {Steinlechner}}]{Krzic2023}%
  \BibitemOpen
  \bibfield  {author} {\bibinfo {author} {\bibfnamefont {A.}~\bibnamefont
  {Kržič}}, \bibinfo {author} {\bibfnamefont {S.}~\bibnamefont {Sharma}},
  \bibinfo {author} {\bibfnamefont {C.}~\bibnamefont {Spiess}}, \bibinfo
  {author} {\bibfnamefont {U.}~\bibnamefont {Chandrashekara}}, \bibinfo
  {author} {\bibfnamefont {S.}~\bibnamefont {Töpfer}}, \bibinfo {author}
  {\bibfnamefont {G.}~\bibnamefont {Sauer}}, \bibinfo {author} {\bibfnamefont
  {L.~J.}\ \bibnamefont {González-Martín~del Campo}}, \bibinfo {author}
  {\bibfnamefont {T.}~\bibnamefont {Kopf}}, \bibinfo {author} {\bibfnamefont
  {S.}~\bibnamefont {Petscharnig}}, \bibinfo {author} {\bibfnamefont
  {T.}~\bibnamefont {Grafenauer}}, \bibinfo {author} {\bibfnamefont
  {R.}~\bibnamefont {Lieger}}, \bibinfo {author} {\bibfnamefont
  {B.}~\bibnamefont {Ömer}}, \bibinfo {author} {\bibfnamefont
  {C.}~\bibnamefont {Pacher}}, \bibinfo {author} {\bibfnamefont
  {R.}~\bibnamefont {Berlich}}, \bibinfo {author} {\bibfnamefont
  {T.}~\bibnamefont {Peschel}}, \bibinfo {author} {\bibfnamefont
  {C.}~\bibnamefont {Damm}}, \bibinfo {author} {\bibfnamefont {S.}~\bibnamefont
  {Risse}}, \bibinfo {author} {\bibfnamefont {M.}~\bibnamefont {Goy}}, \bibinfo
  {author} {\bibfnamefont {D.}~\bibnamefont {Rieländer}}, \bibinfo {author}
  {\bibfnamefont {A.}~\bibnamefont {Tünnermann}},\ and\ \bibinfo {author}
  {\bibfnamefont {F.}~\bibnamefont {Steinlechner}},\ }\bibfield  {title}
  {\bibinfo {title} {Towards metropolitan free-space quantum networks},\ }\href
  {https://doi.org/10.1038/s41534-023-00754-0} {\bibfield  {journal} {\bibinfo
  {journal} {npj Quantum Information}\ }\textbf {\bibinfo {volume} {9}},\
  \bibinfo {pages} {95} (\bibinfo {year} {2023})}\BibitemShut {NoStop}%
\bibitem [{\citenamefont {Zhang}\ \emph {et~al.}(2025)\citenamefont {Zhang},
  \citenamefont {Zhang}, \citenamefont {Eng}, \citenamefont {Meunier},
  \citenamefont {Yang}, \citenamefont {Ling}, \citenamefont {Z\'u\~niga
  P\'erez},\ and\ \citenamefont {Gao}}]{Zhang2025}%
  \BibitemOpen
  \bibfield  {author} {\bibinfo {author} {\bibfnamefont {H.}~\bibnamefont
  {Zhang}}, \bibinfo {author} {\bibfnamefont {X.}~\bibnamefont {Zhang}},
  \bibinfo {author} {\bibfnamefont {J.}~\bibnamefont {Eng}}, \bibinfo {author}
  {\bibfnamefont {M.}~\bibnamefont {Meunier}}, \bibinfo {author} {\bibfnamefont
  {Y.}~\bibnamefont {Yang}}, \bibinfo {author} {\bibfnamefont {A.}~\bibnamefont
  {Ling}}, \bibinfo {author} {\bibfnamefont {J.}~\bibnamefont {Z\'u\~niga
  P\'erez}},\ and\ \bibinfo {author} {\bibfnamefont {W.}~\bibnamefont {Gao}},\
  }\bibfield  {title} {\bibinfo {title} {Metropolitan quantum key distribution
  using a $\mathrm{Ga}\mathrm{N}$-based room-temperature telecommunication
  single-photon source},\ }\href
  {https://doi.org/10.1103/PhysRevApplied.23.054022} {\bibfield  {journal}
  {\bibinfo  {journal} {Phys. Rev. Appl.}\ }\textbf {\bibinfo {volume} {23}},\
  \bibinfo {pages} {054022} (\bibinfo {year} {2025})}\BibitemShut {NoStop}%
\bibitem [{\citenamefont {Scarani}\ \emph {et~al.}(2009)\citenamefont
  {Scarani}, \citenamefont {Bechmann-Pasquinucci}, \citenamefont {Cerf},
  \citenamefont {Du\ifmmode~\check{s}\else \v{s}\fi{}ek}, \citenamefont
  {L\"utkenhaus},\ and\ \citenamefont {Peev}}]{Scarani2009}%
  \BibitemOpen
  \bibfield  {author} {\bibinfo {author} {\bibfnamefont {V.}~\bibnamefont
  {Scarani}}, \bibinfo {author} {\bibfnamefont {H.}~\bibnamefont
  {Bechmann-Pasquinucci}}, \bibinfo {author} {\bibfnamefont {N.~J.}\
  \bibnamefont {Cerf}}, \bibinfo {author} {\bibfnamefont {M.}~\bibnamefont
  {Du\ifmmode~\check{s}\else \v{s}\fi{}ek}}, \bibinfo {author} {\bibfnamefont
  {N.}~\bibnamefont {L\"utkenhaus}},\ and\ \bibinfo {author} {\bibfnamefont
  {M.}~\bibnamefont {Peev}},\ }\bibfield  {title} {\bibinfo {title} {The
  security of practical quantum key distribution},\ }\href
  {https://doi.org/10.1103/RevModPhys.81.1301} {\bibfield  {journal} {\bibinfo
  {journal} {Rev. Mod. Phys.}\ }\textbf {\bibinfo {volume} {81}},\ \bibinfo
  {pages} {1301} (\bibinfo {year} {2009})}\BibitemShut {NoStop}%
\bibitem [{\citenamefont {Jasim}\ \emph {et~al.}(2015)\citenamefont {Jasim},
  \citenamefont {Abbas}, \citenamefont {El-Horbaty},\ and\ \citenamefont
  {Salem}}]{Jasim2015}%
  \BibitemOpen
  \bibfield  {author} {\bibinfo {author} {\bibfnamefont {O.~K.}\ \bibnamefont
  {Jasim}}, \bibinfo {author} {\bibfnamefont {S.}~\bibnamefont {Abbas}},
  \bibinfo {author} {\bibfnamefont {E.-S.~M.}\ \bibnamefont {El-Horbaty}},\
  and\ \bibinfo {author} {\bibfnamefont {A.-B.~M.}\ \bibnamefont {Salem}},\
  }\bibfield  {title} {\bibinfo {title} {Quantum key distribution: Simulation
  and characterizations},\ }\href
  {https://doi.org/https://doi.org/10.1016/j.procs.2015.09.014} {\bibfield
  {journal} {\bibinfo  {journal} {Procedia Computer Science}\ }\textbf
  {\bibinfo {volume} {65}},\ \bibinfo {pages} {701} (\bibinfo {year}
  {2015})}\BibitemShut {NoStop}%
\bibitem [{\citenamefont {Kohnle}\ and\ \citenamefont
  {Rizzoli}(2017)}]{Kohnle2017}%
  \BibitemOpen
  \bibfield  {author} {\bibinfo {author} {\bibfnamefont {A.}~\bibnamefont
  {Kohnle}}\ and\ \bibinfo {author} {\bibfnamefont {A.}~\bibnamefont
  {Rizzoli}},\ }\bibfield  {title} {\bibinfo {title} {Interactive simulations
  for quantum key distribution},\ }\href
  {https://doi.org/10.1088/1361-6404/aa62c8} {\bibfield  {journal} {\bibinfo
  {journal} {European Journal of Physics}\ }\textbf {\bibinfo {volume} {38}},\
  \bibinfo {pages} {035403} (\bibinfo {year} {2017})}\BibitemShut {NoStop}%
\bibitem [{\citenamefont {Wu}\ \emph {et~al.}(2021)\citenamefont {Wu},
  \citenamefont {Kolar}, \citenamefont {Chung}, \citenamefont {Jin},
  \citenamefont {Zhong}, \citenamefont {Kettimuthu},\ and\ \citenamefont
  {Suchara}}]{Wu2021}%
  \BibitemOpen
  \bibfield  {author} {\bibinfo {author} {\bibfnamefont {X.}~\bibnamefont
  {Wu}}, \bibinfo {author} {\bibfnamefont {A.}~\bibnamefont {Kolar}}, \bibinfo
  {author} {\bibfnamefont {J.}~\bibnamefont {Chung}}, \bibinfo {author}
  {\bibfnamefont {D.}~\bibnamefont {Jin}}, \bibinfo {author} {\bibfnamefont
  {T.}~\bibnamefont {Zhong}}, \bibinfo {author} {\bibfnamefont
  {R.}~\bibnamefont {Kettimuthu}},\ and\ \bibinfo {author} {\bibfnamefont
  {M.}~\bibnamefont {Suchara}},\ }\bibfield  {title} {\bibinfo {title}
  {Sequence: a customizable discrete-event simulator of quantum networks},\
  }\href {https://doi.org/10.1088/2058-9565/ac22f6} {\bibfield  {journal}
  {\bibinfo  {journal} {Quantum Science and Technology}\ }\textbf {\bibinfo
  {volume} {6}},\ \bibinfo {pages} {045027} (\bibinfo {year}
  {2021})}\BibitemShut {NoStop}%
\bibitem [{\citenamefont {Semenenko}\ \emph {et~al.}(2022)\citenamefont
  {Semenenko}, \citenamefont {Hu}, \citenamefont {Figueroa},\ and\
  \citenamefont {Perebeinos}}]{Semenenko2022}%
  \BibitemOpen
  \bibfield  {author} {\bibinfo {author} {\bibfnamefont {V.}~\bibnamefont
  {Semenenko}}, \bibinfo {author} {\bibfnamefont {X.}~\bibnamefont {Hu}},
  \bibinfo {author} {\bibfnamefont {E.}~\bibnamefont {Figueroa}},\ and\
  \bibinfo {author} {\bibfnamefont {V.}~\bibnamefont {Perebeinos}},\ }\bibfield
   {title} {\bibinfo {title} {Entanglement generation in a quantum network with
  finite quantum memory lifetime},\ }\href {https://doi.org/10.1116/5.0082239}
  {\bibfield  {journal} {\bibinfo  {journal} {AVS Quantum Science}\ }\textbf
  {\bibinfo {volume} {4}},\ \bibinfo {pages} {012002} (\bibinfo {year}
  {2022})},\ \Eprint {https://arxiv.org/abs/2110.01061} {arXiv:2110.01061}
  \BibitemShut {NoStop}%
\bibitem [{\citenamefont {Ahmed}\ \emph {et~al.}(2026)\citenamefont {Ahmed},
  \citenamefont {Nada}, \citenamefont {Hussain},\ and\ \citenamefont
  {Kupchak}}]{Ahmed2026}%
  \BibitemOpen
  \bibfield  {author} {\bibinfo {author} {\bibfnamefont {T.~R.}\ \bibnamefont
  {Ahmed}}, \bibinfo {author} {\bibfnamefont {F.}~\bibnamefont {Nada}},
  \bibinfo {author} {\bibfnamefont {A.}~\bibnamefont {Hussain}},\ and\ \bibinfo
  {author} {\bibfnamefont {C.}~\bibnamefont {Kupchak}},\ }\bibfield  {title}
  {\bibinfo {title} {Entanglement distribution modelling with quantum memories
  in a global and local clock system},\ }\href
  {https://doi.org/https://doi.org/10.1049/qtc2.70036} {\bibfield  {journal}
  {\bibinfo  {journal} {IET Quantum Communication}\ }\textbf {\bibinfo {volume}
  {7}},\ \bibinfo {pages} {e70036} (\bibinfo {year} {2026})},\ \Eprint
  {https://arxiv.org/abs/https://ietresearch.onlinelibrary.wiley.com/doi/pdf/10.1049/qtc2.70036}
  {https://ietresearch.onlinelibrary.wiley.com/doi/pdf/10.1049/qtc2.70036}
  \BibitemShut {NoStop}%
\bibitem [{\citenamefont {Simon}\ \emph {et~al.}(2018)\citenamefont {Simon},
  \citenamefont {Huff}, \citenamefont {Meier}, \citenamefont {Mailloux},\ and\
  \citenamefont {Harrell}}]{Simon2018}%
  \BibitemOpen
  \bibfield  {author} {\bibinfo {author} {\bibfnamefont {G.~K.}\ \bibnamefont
  {Simon}}, \bibinfo {author} {\bibfnamefont {B.~K.}\ \bibnamefont {Huff}},
  \bibinfo {author} {\bibfnamefont {W.~M.}\ \bibnamefont {Meier}}, \bibinfo
  {author} {\bibfnamefont {L.~O.}\ \bibnamefont {Mailloux}},\ and\ \bibinfo
  {author} {\bibfnamefont {L.~E.}\ \bibnamefont {Harrell}},\ }\bibfield
  {title} {\bibinfo {title} {Quantification of the impact of photon
  distinguishability on measurement-device-independent quantum key
  distribution},\ }\href {https://doi.org/10.3390/electronics7040049}
  {\bibfield  {journal} {\bibinfo  {journal} {Electronics}\ }\textbf {\bibinfo
  {volume} {7}},\ \bibinfo {pages} {49} (\bibinfo {year} {2018})}\BibitemShut
  {NoStop}%
\bibitem [{\citenamefont {Abruzzo}\ \emph {et~al.}(2014)\citenamefont
  {Abruzzo}, \citenamefont {Kampermann},\ and\ \citenamefont
  {Bru\ss{}}}]{Abruzzo2014}%
  \BibitemOpen
  \bibfield  {author} {\bibinfo {author} {\bibfnamefont {S.}~\bibnamefont
  {Abruzzo}}, \bibinfo {author} {\bibfnamefont {H.}~\bibnamefont
  {Kampermann}},\ and\ \bibinfo {author} {\bibfnamefont {D.}~\bibnamefont
  {Bru\ss{}}},\ }\bibfield  {title} {\bibinfo {title}
  {Measurement-device-independent quantum key distribution with quantum
  memories},\ }\href {https://doi.org/10.1103/PhysRevA.89.012301} {\bibfield
  {journal} {\bibinfo  {journal} {Phys. Rev. A}\ }\textbf {\bibinfo {volume}
  {89}},\ \bibinfo {pages} {012301} (\bibinfo {year} {2014})}\BibitemShut
  {NoStop}%
\bibitem [{\citenamefont {Chelluri}\ \emph {et~al.}(2026)\citenamefont
  {Chelluri}, \citenamefont {Sharma}, \citenamefont {Schmidt}, \citenamefont
  {Kusminskiy},\ and\ \citenamefont {van Loock}}]{Chelluri2026}%
  \BibitemOpen
  \bibfield  {author} {\bibinfo {author} {\bibfnamefont {S.~S.}\ \bibnamefont
  {Chelluri}}, \bibinfo {author} {\bibfnamefont {S.}~\bibnamefont {Sharma}},
  \bibinfo {author} {\bibfnamefont {F.}~\bibnamefont {Schmidt}}, \bibinfo
  {author} {\bibfnamefont {S.~V.}\ \bibnamefont {Kusminskiy}},\ and\ \bibinfo
  {author} {\bibfnamefont {P.}~\bibnamefont {van Loock}},\ }\bibfield  {title}
  {\bibinfo {title} {Bosonic quantum error correction with microwave cavities
  for quantum repeaters},\ }\href {https://doi.org/10.1103/hkvx-rks6}
  {\bibfield  {journal} {\bibinfo  {journal} {Phys. Rev. A}\ }\textbf {\bibinfo
  {volume} {114}},\ \bibinfo {pages} {012616} (\bibinfo {year}
  {2026})}\BibitemShut {NoStop}%
\bibitem [{\citenamefont {MacRae}\ \emph {et~al.}(2012)\citenamefont {MacRae},
  \citenamefont {Brannan}, \citenamefont {Achal},\ and\ \citenamefont
  {Lvovsky}}]{Macrae2012}%
  \BibitemOpen
  \bibfield  {author} {\bibinfo {author} {\bibfnamefont {A.}~\bibnamefont
  {MacRae}}, \bibinfo {author} {\bibfnamefont {T.}~\bibnamefont {Brannan}},
  \bibinfo {author} {\bibfnamefont {R.}~\bibnamefont {Achal}},\ and\ \bibinfo
  {author} {\bibfnamefont {A.~I.}\ \bibnamefont {Lvovsky}},\ }\bibfield
  {title} {\bibinfo {title} {Tomography of a high-purity narrowband photon from
  a transient atomic collective excitation},\ }\href
  {https://doi.org/10.1103/PhysRevLett.109.033601} {\bibfield  {journal}
  {\bibinfo  {journal} {Phys. Rev. Lett.}\ }\textbf {\bibinfo {volume} {109}},\
  \bibinfo {pages} {033601} (\bibinfo {year} {2012})}\BibitemShut {NoStop}%
\bibitem [{\citenamefont {Zhan}\ \emph {et~al.}(2023)\citenamefont {Zhan} \emph
  {et~al.}}]{Zhan2023}%
  \BibitemOpen
  \bibfield  {author} {\bibinfo {author} {\bibfnamefont {X.-H.}\ \bibnamefont
  {Zhan}} \emph {et~al.},\ }\bibfield  {title} {\bibinfo {title}
  {Measurement-device-independent quantum key distribution with practical
  spontaneous parametric down-conversion sources},\ }\href
  {https://doi.org/10.1103/PhysRevApplied.20.034069} {\bibfield  {journal}
  {\bibinfo  {journal} {Phys. Rev. Applied}\ }\textbf {\bibinfo {volume}
  {20}},\ \bibinfo {pages} {034069} (\bibinfo {year} {2023})}\BibitemShut
  {NoStop}%
\bibitem [{\citenamefont {Chen}\ \emph {et~al.}(2026)\citenamefont {Chen},
  \citenamefont {Zhong}, \citenamefont {Zhan}, \citenamefont {Wang},
  \citenamefont {Yin}, \citenamefont {He}, \citenamefont {Chen}, \citenamefont
  {Guo},\ and\ \citenamefont {Han}}]{Chen2026}%
  \BibitemOpen
  \bibfield  {author} {\bibinfo {author} {\bibfnamefont {H.-Y.}\ \bibnamefont
  {Chen}}, \bibinfo {author} {\bibfnamefont {Z.-Q.}\ \bibnamefont {Zhong}},
  \bibinfo {author} {\bibfnamefont {X.-H.}\ \bibnamefont {Zhan}}, \bibinfo
  {author} {\bibfnamefont {S.}~\bibnamefont {Wang}}, \bibinfo {author}
  {\bibfnamefont {Z.-Q.}\ \bibnamefont {Yin}}, \bibinfo {author} {\bibfnamefont
  {D.-Y.}\ \bibnamefont {He}}, \bibinfo {author} {\bibfnamefont
  {W.}~\bibnamefont {Chen}}, \bibinfo {author} {\bibfnamefont {G.-C.}\
  \bibnamefont {Guo}},\ and\ \bibinfo {author} {\bibfnamefont {Z.-F.}\
  \bibnamefont {Han}},\ }\bibfield  {title} {\bibinfo {title}
  {Measurement-device-independent quantum key distribution based on
  postprocessing time-domain filtering},\ }\href
  {https://doi.org/10.1103/gvvj-17xy} {\bibfield  {journal} {\bibinfo
  {journal} {Phys. Rev. Applied}\ }\textbf {\bibinfo {volume} {25}},\ \bibinfo
  {pages} {054032} (\bibinfo {year} {2026})}\BibitemShut {NoStop}%
\bibitem [{\citenamefont {Lütkenhaus}\ and\ \citenamefont
  {Jahma}(2002)}]{Lutkenhaus2002}%
  \BibitemOpen
  \bibfield  {author} {\bibinfo {author} {\bibfnamefont {N.}~\bibnamefont
  {Lütkenhaus}}\ and\ \bibinfo {author} {\bibfnamefont {M.}~\bibnamefont
  {Jahma}},\ }\bibfield  {title} {\bibinfo {title} {Quantum key distribution
  with realistic states: photon-number statistics in the photon-number
  splitting attack},\ }\href {https://doi.org/10.1088/1367-2630/4/1/344}
  {\bibfield  {journal} {\bibinfo  {journal} {New Journal of Physics}\ }\textbf
  {\bibinfo {volume} {4}},\ \bibinfo {pages} {44} (\bibinfo {year}
  {2002})}\BibitemShut {NoStop}%
\bibitem [{\citenamefont {Ma}\ \emph {et~al.}(2005)\citenamefont {Ma},
  \citenamefont {Qi}, \citenamefont {Zhao},\ and\ \citenamefont
  {Lo}}]{LoDecoyStates2005}%
  \BibitemOpen
  \bibfield  {author} {\bibinfo {author} {\bibfnamefont {X.}~\bibnamefont
  {Ma}}, \bibinfo {author} {\bibfnamefont {B.}~\bibnamefont {Qi}}, \bibinfo
  {author} {\bibfnamefont {Y.}~\bibnamefont {Zhao}},\ and\ \bibinfo {author}
  {\bibfnamefont {H.-K.}\ \bibnamefont {Lo}},\ }\bibfield  {title} {\bibinfo
  {title} {Practical decoy state for quantum key distribution},\ }\bibfield
  {journal} {\bibinfo  {journal} {Physical Review A}\ }\textbf {\bibinfo
  {volume} {72}},\ \href {https://doi.org/10.1103/physreva.72.012326}
  {10.1103/physreva.72.012326} (\bibinfo {year} {2005})\BibitemShut {NoStop}%
\bibitem [{\citenamefont {Wang}\ \emph
  {et~al.}(2022{\natexlab{b}})\citenamefont {Wang}, \citenamefont {Craddock},
  \citenamefont {Sekelsky}, \citenamefont {Flament},\ and\ \citenamefont
  {Namazi}}]{Wang2022}%
  \BibitemOpen
  \bibfield  {author} {\bibinfo {author} {\bibfnamefont {Y.}~\bibnamefont
  {Wang}}, \bibinfo {author} {\bibfnamefont {A.~N.}\ \bibnamefont {Craddock}},
  \bibinfo {author} {\bibfnamefont {R.}~\bibnamefont {Sekelsky}}, \bibinfo
  {author} {\bibfnamefont {M.}~\bibnamefont {Flament}},\ and\ \bibinfo {author}
  {\bibfnamefont {M.}~\bibnamefont {Namazi}},\ }\bibfield  {title} {\bibinfo
  {title} {Field-deployable quantum memory for quantum networking},\ }\href
  {https://doi.org/10.1103/PhysRevApplied.18.044058} {\bibfield  {journal}
  {\bibinfo  {journal} {Phys. Rev. Appl.}\ }\textbf {\bibinfo {volume} {18}},\
  \bibinfo {pages} {044058} (\bibinfo {year} {2022}{\natexlab{b}})}\BibitemShut
  {NoStop}%
\bibitem [{\citenamefont {Wang}\ \emph {et~al.}(2025)\citenamefont {Wang},
  \citenamefont {Craddock}, \citenamefont {Mendoza}, \citenamefont {Sekelsky},
  \citenamefont {Flament},\ and\ \citenamefont {Namazi}}]{Wang2025}%
  \BibitemOpen
  \bibfield  {author} {\bibinfo {author} {\bibfnamefont {Y.}~\bibnamefont
  {Wang}}, \bibinfo {author} {\bibfnamefont {A.~N.}\ \bibnamefont {Craddock}},
  \bibinfo {author} {\bibfnamefont {J.~M.}\ \bibnamefont {Mendoza}}, \bibinfo
  {author} {\bibfnamefont {R.}~\bibnamefont {Sekelsky}}, \bibinfo {author}
  {\bibfnamefont {M.}~\bibnamefont {Flament}},\ and\ \bibinfo {author}
  {\bibfnamefont {M.}~\bibnamefont {Namazi}},\ }\bibfield  {title} {\bibinfo
  {title} {High-fidelity entanglement between a telecom photon and a
  room-temperature quantum memory},\ }\href {https://arxiv.org/abs/2503.11564}
  {\bibfield  {journal} {\bibinfo  {journal} {arXiv preprint arXiv:2503.11564}\
  } (\bibinfo {year} {2025})}\BibitemShut {NoStop}%
\bibitem [{\citenamefont {Guo}\ \emph {et~al.}(2025)\citenamefont {Guo},
  \citenamefont {Wu}, \citenamefont {Bao}, \citenamefont {Yang}, \citenamefont
  {Wu}, \citenamefont {Chen},\ and\ \citenamefont {Zhang}}]{Guo2025}%
  \BibitemOpen
  \bibfield  {author} {\bibinfo {author} {\bibfnamefont {J.}~\bibnamefont
  {Guo}}, \bibinfo {author} {\bibfnamefont {Z.}~\bibnamefont {Wu}}, \bibinfo
  {author} {\bibfnamefont {G.}~\bibnamefont {Bao}}, \bibinfo {author}
  {\bibfnamefont {P.}~\bibnamefont {Yang}}, \bibinfo {author} {\bibfnamefont
  {Y.}~\bibnamefont {Wu}}, \bibinfo {author} {\bibfnamefont {L.~Q.}\
  \bibnamefont {Chen}},\ and\ \bibinfo {author} {\bibfnamefont
  {W.}~\bibnamefont {Zhang}},\ }\bibfield  {title} {\bibinfo {title}
  {Near-perfect broadband quantum memory enabled by intelligent spin-wave
  compaction},\ }\href {https://doi.org/10.1103/kbwj-md9n} {\bibfield
  {journal} {\bibinfo  {journal} {Phys. Rev. Lett.}\ }\textbf {\bibinfo
  {volume} {135}},\ \bibinfo {pages} {170802} (\bibinfo {year}
  {2025})}\BibitemShut {NoStop}%
\bibitem [{\citenamefont {Finkelstein}\ \emph {et~al.}(2018)\citenamefont
  {Finkelstein}, \citenamefont {Poem}, \citenamefont {Michel}, \citenamefont
  {Lahad},\ and\ \citenamefont {Firstenberg}}]{Finkelstein2018}%
  \BibitemOpen
  \bibfield  {author} {\bibinfo {author} {\bibfnamefont {R.}~\bibnamefont
  {Finkelstein}}, \bibinfo {author} {\bibfnamefont {E.}~\bibnamefont {Poem}},
  \bibinfo {author} {\bibfnamefont {O.}~\bibnamefont {Michel}}, \bibinfo
  {author} {\bibfnamefont {O.}~\bibnamefont {Lahad}},\ and\ \bibinfo {author}
  {\bibfnamefont {O.}~\bibnamefont {Firstenberg}},\ }\bibfield  {title}
  {\bibinfo {title} {Fast, noise-free memory for photon synchronization at room
  temperature},\ }\href {https://doi.org/10.1126/sciadv.aap8598} {\bibfield
  {journal} {\bibinfo  {journal} {Science Advances}\ }\textbf {\bibinfo
  {volume} {4}},\ \bibinfo {pages} {eaap8598} (\bibinfo {year} {2018})},\
  \Eprint
  {https://arxiv.org/abs/https://www.science.org/doi/pdf/10.1126/sciadv.aap8598}
  {https://www.science.org/doi/pdf/10.1126/sciadv.aap8598} \BibitemShut
  {NoStop}%
\bibitem [{\citenamefont {Michelberger}\ \emph {et~al.}(2015)\citenamefont
  {Michelberger}, \citenamefont {Champion}, \citenamefont {Sprague},
  \citenamefont {Kaczmarek}, \citenamefont {Barbieri}, \citenamefont {Jin},
  \citenamefont {England}, \citenamefont {Kolthammer}, \citenamefont
  {Saunders}, \citenamefont {Nunn},\ and\ \citenamefont
  {Walmsley}}]{Michelberger2015}%
  \BibitemOpen
  \bibfield  {author} {\bibinfo {author} {\bibfnamefont {P.~S.}\ \bibnamefont
  {Michelberger}}, \bibinfo {author} {\bibfnamefont {T.~F.~M.}\ \bibnamefont
  {Champion}}, \bibinfo {author} {\bibfnamefont {M.~R.}\ \bibnamefont
  {Sprague}}, \bibinfo {author} {\bibfnamefont {K.~T.}\ \bibnamefont
  {Kaczmarek}}, \bibinfo {author} {\bibfnamefont {M.}~\bibnamefont {Barbieri}},
  \bibinfo {author} {\bibfnamefont {X.~M.}\ \bibnamefont {Jin}}, \bibinfo
  {author} {\bibfnamefont {D.~G.}\ \bibnamefont {England}}, \bibinfo {author}
  {\bibfnamefont {W.~S.}\ \bibnamefont {Kolthammer}}, \bibinfo {author}
  {\bibfnamefont {D.~J.}\ \bibnamefont {Saunders}}, \bibinfo {author}
  {\bibfnamefont {J.}~\bibnamefont {Nunn}},\ and\ \bibinfo {author}
  {\bibfnamefont {I.~A.}\ \bibnamefont {Walmsley}},\ }\bibfield  {title}
  {\bibinfo {title} {Interfacing ghz-bandwidth heralded single photons with a
  warm vapour raman memory},\ }\href
  {https://doi.org/10.1088/1367-2630/17/4/043006} {\bibfield  {journal}
  {\bibinfo  {journal} {New Journal of Physics}\ }\textbf {\bibinfo {volume}
  {17}},\ \bibinfo {pages} {043006} (\bibinfo {year} {2015})}\BibitemShut
  {NoStop}%
\bibitem [{\citenamefont {Tsai}\ \emph {et~al.}(2020)\citenamefont {Tsai},
  \citenamefont {Hsiao},\ and\ \citenamefont {Chen}}]{Cheng2020}%
  \BibitemOpen
  \bibfield  {author} {\bibinfo {author} {\bibfnamefont {P.-J.}\ \bibnamefont
  {Tsai}}, \bibinfo {author} {\bibfnamefont {Y.-F.}\ \bibnamefont {Hsiao}},\
  and\ \bibinfo {author} {\bibfnamefont {Y.-C.}\ \bibnamefont {Chen}},\
  }\bibfield  {title} {\bibinfo {title} {Quantum storage and manipulation of
  heralded single photons in atomic memories based on electromagnetically
  induced transparency},\ }\href
  {https://doi.org/10.1103/PhysRevResearch.2.033155} {\bibfield  {journal}
  {\bibinfo  {journal} {Phys. Rev. Res.}\ }\textbf {\bibinfo {volume} {2}},\
  \bibinfo {pages} {033155} (\bibinfo {year} {2020})}\BibitemShut {NoStop}%
\bibitem [{\citenamefont {Saleh}\ and\ \citenamefont
  {Teich}(2007)}]{SalehTeich2007}%
  \BibitemOpen
  \bibfield  {author} {\bibinfo {author} {\bibfnamefont {B.~E.~A.}\
  \bibnamefont {Saleh}}\ and\ \bibinfo {author} {\bibfnamefont {M.~C.}\
  \bibnamefont {Teich}},\ }\href@noop {} {\emph {\bibinfo {title} {Fundamentals
  of Photonics}}},\ \bibinfo {edition} {2nd}\ ed.\ (\bibinfo  {publisher}
  {Wiley},\ \bibinfo {address} {Hoboken, NJ},\ \bibinfo {year}
  {2007})\BibitemShut {NoStop}%
\bibitem [{\citenamefont {Tan}\ \emph {et~al.}(2024)\citenamefont {Tan},
  \citenamefont {Wang}, \citenamefont {Wu},\ and\ \citenamefont
  {He}}]{Tan2024}%
  \BibitemOpen
  \bibfield  {author} {\bibinfo {author} {\bibfnamefont {Y.}~\bibnamefont
  {Tan}}, \bibinfo {author} {\bibfnamefont {J.}~\bibnamefont {Wang}}, \bibinfo
  {author} {\bibfnamefont {J.}~\bibnamefont {Wu}},\ and\ \bibinfo {author}
  {\bibfnamefont {Z.}~\bibnamefont {He}},\ }\bibfield  {title} {\bibinfo
  {title} {Real-time polarization compensation method in quantum communication
  based on channel muller parameters detection},\ }\href
  {https://doi.org/10.1038/s44172-024-00198-0} {\bibfield  {journal} {\bibinfo
  {journal} {Communications Engineering}\ }\textbf {\bibinfo {volume} {3}},\
  \bibinfo {pages} {57} (\bibinfo {year} {2024})}\BibitemShut {NoStop}%
\bibitem [{\citenamefont {Xu}\ \emph {et~al.}(2013)\citenamefont {Xu},
  \citenamefont {Curty}, \citenamefont {Qi},\ and\ \citenamefont
  {Lo}}]{ZhongLo2015}%
  \BibitemOpen
  \bibfield  {author} {\bibinfo {author} {\bibfnamefont {F.}~\bibnamefont
  {Xu}}, \bibinfo {author} {\bibfnamefont {M.}~\bibnamefont {Curty}}, \bibinfo
  {author} {\bibfnamefont {B.}~\bibnamefont {Qi}},\ and\ \bibinfo {author}
  {\bibfnamefont {H.-K.}\ \bibnamefont {Lo}},\ }\bibfield  {title} {\bibinfo
  {title} {Practical aspects of measurement-device-independent quantum key
  distribution},\ }\href {https://doi.org/10.1088/1367-2630/15/11/113007}
  {\bibfield  {journal} {\bibinfo  {journal} {New Journal of Physics}\ }\textbf
  {\bibinfo {volume} {15}},\ \bibinfo {pages} {113007} (\bibinfo {year}
  {2013})}\BibitemShut {NoStop}%
\bibitem [{\citenamefont {Li}\ \emph {et~al.}(2023)\citenamefont {Li},
  \citenamefont {Fan-Yuan}, \citenamefont {Wang}, \citenamefont {Wang},
  \citenamefont {Lu} \emph {et~al.}}]{Li2023}%
  \BibitemOpen
  \bibfield  {author} {\bibinfo {author} {\bibfnamefont {J.-X.}\ \bibnamefont
  {Li}}, \bibinfo {author} {\bibfnamefont {G.-J.}\ \bibnamefont {Fan-Yuan}},
  \bibinfo {author} {\bibfnamefont {S.}~\bibnamefont {Wang}}, \bibinfo {author}
  {\bibfnamefont {Z.-H.}\ \bibnamefont {Wang}}, \bibinfo {author}
  {\bibfnamefont {F.-Y.}\ \bibnamefont {Lu}}, \emph {et~al.},\ }\bibfield
  {title} {\bibinfo {title} {Polarization alignment in
  measurement-device-independent quantum key distribution with intrinsic
  events},\ }\href {https://doi.org/10.1103/PhysRevApplied.20.054041}
  {\bibfield  {journal} {\bibinfo  {journal} {Phys. Rev. Applied}\ }\textbf
  {\bibinfo {volume} {20}},\ \bibinfo {pages} {054041} (\bibinfo {year}
  {2023})}\BibitemShut {NoStop}%
\bibitem [{\citenamefont {Yastremski}\ \emph {et~al.}(2025)\citenamefont
  {Yastremski}, \citenamefont {Godin},\ and\ \citenamefont
  {et~al.}}]{Yastremski2025}%
  \BibitemOpen
  \bibfield  {author} {\bibinfo {author} {\bibfnamefont {M.}~\bibnamefont
  {Yastremski}}, \bibinfo {author} {\bibfnamefont {P.~J.}\ \bibnamefont
  {Godin}},\ and\ \bibinfo {author} {\bibnamefont {et~al.}},\ }\bibfield
  {title} {\bibinfo {title} {Estimating the impact of light pollution on
  quantum communication between {QEYSSat} and canadian quantum ground station
  sites},\ }\bibfield  {journal} {\bibinfo  {journal} {EPJ Quantum Technology}\
  }\textbf {\bibinfo {volume} {12}},\ \href
  {https://doi.org/10.1140/epjqt/s40507-025-00331-8}
  {10.1140/epjqt/s40507-025-00331-8} (\bibinfo {year} {2025})\BibitemShut
  {NoStop}%
\bibitem [{\citenamefont {Bebrov}(2024)}]{Bebrov2024}%
  \BibitemOpen
  \bibfield  {author} {\bibinfo {author} {\bibfnamefont {G.}~\bibnamefont
  {Bebrov}},\ }\bibfield  {title} {\bibinfo {title} {On the (relation between)
  efficiency and secret key rate of qkd},\ }\href
  {https://doi.org/10.1038/s41598-024-54246-y} {\bibfield  {journal} {\bibinfo
  {journal} {Scientific Reports}\ }\textbf {\bibinfo {volume} {14}},\ \bibinfo
  {pages} {3638} (\bibinfo {year} {2024})}\BibitemShut {NoStop}%
\bibitem [{\citenamefont {Mo}\ \emph {et~al.}(2011)\citenamefont {Mo},
  \citenamefont {Lucio-Martinez}, \citenamefont {Chan}, \citenamefont {Healey},
  \citenamefont {Hosier},\ and\ \citenamefont {Tittel}}]{Mo2011}%
  \BibitemOpen
  \bibfield  {author} {\bibinfo {author} {\bibfnamefont {X.~F.}\ \bibnamefont
  {Mo}}, \bibinfo {author} {\bibfnamefont {I.}~\bibnamefont {Lucio-Martinez}},
  \bibinfo {author} {\bibfnamefont {P.}~\bibnamefont {Chan}}, \bibinfo {author}
  {\bibfnamefont {C.}~\bibnamefont {Healey}}, \bibinfo {author} {\bibfnamefont
  {S.}~\bibnamefont {Hosier}},\ and\ \bibinfo {author} {\bibfnamefont
  {W.}~\bibnamefont {Tittel}},\ }\bibfield  {title} {\bibinfo {title}
  {Time-cost analysis of a quantum key distribution system clocked at 100
  mhz},\ }\href {https://doi.org/10.1364/OE.19.017729} {\bibfield  {journal}
  {\bibinfo  {journal} {Opt. Express}\ }\textbf {\bibinfo {volume} {19}},\
  \bibinfo {pages} {17729} (\bibinfo {year} {2011})}\BibitemShut {NoStop}%
\bibitem [{\citenamefont {Fan-Yuan}\ \emph {et~al.}(2026)\citenamefont
  {Fan-Yuan}, \citenamefont {Xie}, \citenamefont {He}, \citenamefont {Huang},
  \citenamefont {Wang}, \citenamefont {Yin}, \citenamefont {Chen},
  \citenamefont {Li}, \citenamefont {Mao}, \citenamefont {Zhang}, \citenamefont
  {Wang}, \citenamefont {Guo},\ and\ \citenamefont {Han}}]{FanYuan2026}%
  \BibitemOpen
  \bibfield  {author} {\bibinfo {author} {\bibfnamefont {G.-J.}\ \bibnamefont
  {Fan-Yuan}}, \bibinfo {author} {\bibfnamefont {W.-X.}\ \bibnamefont {Xie}},
  \bibinfo {author} {\bibfnamefont {D.-Y.}\ \bibnamefont {He}}, \bibinfo
  {author} {\bibfnamefont {X.-J.}\ \bibnamefont {Huang}}, \bibinfo {author}
  {\bibfnamefont {S.}~\bibnamefont {Wang}}, \bibinfo {author} {\bibfnamefont
  {Z.-Q.}\ \bibnamefont {Yin}}, \bibinfo {author} {\bibfnamefont
  {W.}~\bibnamefont {Chen}}, \bibinfo {author} {\bibfnamefont {Q.}~\bibnamefont
  {Li}}, \bibinfo {author} {\bibfnamefont {H.-K.}\ \bibnamefont {Mao}},
  \bibinfo {author} {\bibfnamefont {G.-W.}\ \bibnamefont {Zhang}}, \bibinfo
  {author} {\bibfnamefont {Y.-L.}\ \bibnamefont {Wang}}, \bibinfo {author}
  {\bibfnamefont {G.-C.}\ \bibnamefont {Guo}},\ and\ \bibinfo {author}
  {\bibfnamefont {Z.-F.}\ \bibnamefont {Han}},\ }\bibfield  {title} {\bibinfo
  {title} {High-rate quantum key distribution with compact state preparation
  and detection},\ }\href {https://doi.org/10.1073/pnas.2521590123} {\bibfield
  {journal} {\bibinfo  {journal} {Proc. Natl. Acad. Sci. U.S.A.}\ }\textbf
  {\bibinfo {volume} {123}},\ \bibinfo {pages} {e2521590123} (\bibinfo {year}
  {2026})}\BibitemShut {NoStop}%
\bibitem [{\citenamefont {Fan-Yuan}\ \emph {et~al.}(2022)\citenamefont
  {Fan-Yuan}, \citenamefont {Lu}, \citenamefont {Wang}, \citenamefont {Yin},
  \citenamefont {He}, \citenamefont {Chen}, \citenamefont {Zhou}, \citenamefont
  {Wang}, \citenamefont {Teng}, \citenamefont {Guo},\ and\ \citenamefont
  {Han}}]{FanYuan2022}%
  \BibitemOpen
  \bibfield  {author} {\bibinfo {author} {\bibfnamefont {G.-J.}\ \bibnamefont
  {Fan-Yuan}}, \bibinfo {author} {\bibfnamefont {F.-Y.}\ \bibnamefont {Lu}},
  \bibinfo {author} {\bibfnamefont {S.}~\bibnamefont {Wang}}, \bibinfo {author}
  {\bibfnamefont {Z.-Q.}\ \bibnamefont {Yin}}, \bibinfo {author} {\bibfnamefont
  {D.-Y.}\ \bibnamefont {He}}, \bibinfo {author} {\bibfnamefont
  {W.}~\bibnamefont {Chen}}, \bibinfo {author} {\bibfnamefont {Z.}~\bibnamefont
  {Zhou}}, \bibinfo {author} {\bibfnamefont {Z.-H.}\ \bibnamefont {Wang}},
  \bibinfo {author} {\bibfnamefont {J.}~\bibnamefont {Teng}}, \bibinfo {author}
  {\bibfnamefont {G.-C.}\ \bibnamefont {Guo}},\ and\ \bibinfo {author}
  {\bibfnamefont {Z.-F.}\ \bibnamefont {Han}},\ }\bibfield  {title} {\bibinfo
  {title} {Robust and adaptable quantum key distribution network without
  trusted nodes},\ }\href@noop {} {\bibfield  {journal} {\bibinfo  {journal}
  {Optica}\ }\textbf {\bibinfo {volume} {9}},\ \bibinfo {pages} {812} (\bibinfo
  {year} {2022})}\BibitemShut {NoStop}%
\bibitem [{\citenamefont {Lu}\ \emph {et~al.}(2025)\citenamefont {Lu},
  \citenamefont {Wang}, \citenamefont {Zhou}, \citenamefont {Fan},
  \citenamefont {Wang}, \citenamefont {Yin}, \citenamefont {Li}, \citenamefont
  {He}, \citenamefont {Wang}, \citenamefont {Chen}, \citenamefont {Xue},
  \citenamefont {Guo},\ and\ \citenamefont {Han}}]{Lu2025}%
  \BibitemOpen
  \bibfield  {author} {\bibinfo {author} {\bibfnamefont {F.-Y.}\ \bibnamefont
  {Lu}}, \bibinfo {author} {\bibfnamefont {Z.-H.}\ \bibnamefont {Wang}},
  \bibinfo {author} {\bibfnamefont {Y.}~\bibnamefont {Zhou}}, \bibinfo {author}
  {\bibfnamefont {Y.-X.}\ \bibnamefont {Fan}}, \bibinfo {author} {\bibfnamefont
  {S.}~\bibnamefont {Wang}}, \bibinfo {author} {\bibfnamefont {Z.-Q.}\
  \bibnamefont {Yin}}, \bibinfo {author} {\bibfnamefont {J.}~\bibnamefont
  {Li}}, \bibinfo {author} {\bibfnamefont {D.-Y.}\ \bibnamefont {He}}, \bibinfo
  {author} {\bibfnamefont {F.-X.}\ \bibnamefont {Wang}}, \bibinfo {author}
  {\bibfnamefont {W.}~\bibnamefont {Chen}}, \bibinfo {author} {\bibfnamefont
  {K.}~\bibnamefont {Xue}}, \bibinfo {author} {\bibfnamefont {G.-C.}\
  \bibnamefont {Guo}},\ and\ \bibinfo {author} {\bibfnamefont {Z.-F.}\
  \bibnamefont {Han}},\ }\bibfield  {title} {\bibinfo {title} {Fully
  heterogeneous prepare-and-measure quantum network for the next stage of
  quantum internet},\ }\href {https://doi.org/10.1038/s41467-025-66333-3}
  {\bibfield  {journal} {\bibinfo  {journal} {Nature Communications}\ }\textbf
  {\bibinfo {volume} {16}},\ \bibinfo {pages} {11487} (\bibinfo {year}
  {2025})}\BibitemShut {NoStop}%
\end{thebibliography}%

\end{document}